\documentclass[12pt]{article}
\usepackage{epsfig}
\usepackage{amsmath}
\usepackage{amssymb}
\usepackage{graphicx}

\usepackage {subfigure}
\usepackage{hyperref, cite}
\def\be{\begin{equation}}
\def\ee{\end{equation}}
\def\ba{\begin{array}}
\def\ea{\end{array}}
\def\beqn{\begin{eqnarray}}
\def\eeqn{\end{eqnarray}}

\def\bt{\begin{tabular}}
\def\et{\end{tabular}}
\def\bc{\begin{center}}
\def\ec{\end{center}}
\usepackage{epstopdf}
\usepackage{lscape}
\begin{document}

\title{Texture One Zero Dirac Neutrino Mass Matrix With Vanishing Determinant or Trace Condition}
\author{Madan Singh$^{*} $\\
\it Department of Physics, National Institute of Technology Kurukshetra,\\
\it Haryana,136119, India.\\
\it $^{*}$singhmadan179@gmail.com
}
\maketitle

\begin{abstract}
In the light of non-zero and relatively large value of rector mixing angle ($\theta_{13}$), we have performed a detailed analysis of texture one zero neutrino mass matrix $M_{\nu}$ in the scenario of vanishing determinant/trace conditions , assuming the Dirac nature of neutrinos.  In both the scenarios, normal mass ordering is ruled out for all the six possibilities of $M_{\nu}$, however for inverted mass ordering, only two are found to be viable with the current neutrino oscillation data at $3\sigma$ confidence level.  Numerical and some approximate analytical results are presented.
\end{abstract}

\section{Introduction}

The Double Chooz, Daya Bay and RENO Collaborations \cite{1,2,3,4,5,6,7} have finally established the non-zero and relatively large reactor mixing angle $\theta_{13}$, therefore the number of precisely known neutrino oscillation parameters becomes five comprising two mass squared differences ($\delta m^{2}$, $\Delta m^{2}$) and three neutrino mixing angles ($\theta_{12}$, $\theta_{23}$, $\theta_{13}$). However, any general $3\times 3$ neutrino mass matrix contains more parameters than can be measured in realistic experiments.

Several phenomenological schemes in particular, texture zeros \cite{8,9,10,11,12,13, 14,15} have been adopted in the literature in both flavor and non flavor basis, which not only  allows to reduce the number of free parameters of $M_{\nu}$, but also helps to establish some interesting relations between flavor mixing angles and fermion mass ratios \cite{9}. Specifically, in the flavor basis wherein the charged lepton
mass matrix is considered to be diagonal, a particular attention has been paid to explore the viability of texture zero mass matrices for Dirac \cite{13, 14} as well as Majorana \cite{8,9,10,11,12,15} neutrinos 
with the experimental data. In Refs. \cite{8,9,10,11,12,13, 14,15}, most of the texture zero analyses have been carried out assuming  the Majorana nature of neutrinos, because various see-saw mechanisms for neutrino mass generation lead to light Majorana neutrinos. However, considering the present ambiguity on neutrino mass, neutrinos could still be a Dirac particle. The  highly-suppressed Yukawa couplings for Dirac neutrinos can naturally be achieved in the several models with extra spatial dimensions \cite{16} or through radiative mechanisms \cite{17} and also in supersymmetry models \cite{18},  supergravity models \cite{19} of Dirac neutrino masses. Moreover, a common argument in favor of Majorana neutrinos is that the implied lepton number violation can be used to generate the baryon asymmetry of the universe via the leptogenesis mechanism \cite{20}. However, similar argument can be made even for Dirac neutrinos  \cite{21,22}.

Seeking the motivation for Dirac neutrinos from these theoretical grounds, Liu and Zhou \cite{14} have  carried out an analysis of texture zero mass matrices in the flavor basis, and found that all the six possiblities carrying one texture zero in the neutrino mass matrix are experimentally viable. This is not surprising as texture one zero makes available larger parametric space for viability with the data compared with texture two zero case. However, to impart predictability to texture one zero, additional constraints in the form of Det $M_{\nu}$=0 or Tr $M_{\nu}$=0 can be incorporated.  The Det $M_{\nu} = 0$ \cite{23}  condition  can be motivated on various theoretical grounds \cite{24, 25}. The condition Det $M_{\nu} = 0$ is equivalent to assuming one of the neutrinos to be massless. This is realized,
for instance, in the Affleck-Dine scenario for leptogenesis \cite{26} which requires the lightest
neutrino to be practically massless ($m \simeq 10^{-10}$eV) \cite{27, 28}. In Refs. \cite{15, 29}, the implication for the same have been rigoursly studied for texture one zero mass Majorana matrices. The motivation for Tr $M_{\nu}$ = 0 condition, was first put forward in \cite{30} applying a three neutrino framework that simultaneously explains the anomalies of solar and atmospheric neutrino oscillation experiments as well as the LSND experiment. In \cite{31}, X. G. He and  A. Zee  have investigated the CP conserving traceless $M_{\nu}$ for the more realistic case of explaining only the solar neutrino atmospheric and deficits. Further motivation of traceless mass matrices can be provided by models wherein $M_{\nu}$ is constructed through a commutator of two matrices, as it happens in models of radiative mass generation \cite{32}. H. A. Alhendi et.al. \cite{33}  have incorporated the tracless condition with two 2 $\times$ 2 sub-matrices of Majorana mass matrix in the flavor basis and carried out a detailed numerical analysis at 3$\sigma$ confidence level. Also the phenomenological implications of traceless $M_{\nu}$ on neutrino masses, CP violating phases and effective neutrino mass term is studied in Ref.\cite{34}, for both normal and inverted mass ordering and in case of CP conservation and violation. 

Without loss of generality, we consider a neutrino mass matrix $M_{\nu}$ for Dirac neutrinos to be Hermitian by redefining the right-handed neutrino fields. As $M_{\nu}$ is Hermitian, three independent off-diagonal matrix elements are in general complex, while three independent diagonal ones are real.  Following Ref.\cite{14}, the six possible texture one zero hermitian matrices are given in Table \ref{tab1}. The nomenclature is similar to texture one zero for Majorana neutrino except that here neutrino mass matrix is hermitian.
\begin{table}[htp]
\begin{center}
\begin{tabular}{|c|c|c|}
\hline $P_1$ & $P_2$ & $P_3$  \\
\hline $\left(
\begin{array}{ccc}
0 & \Delta & \Delta \\
\Delta^{*} & \times &\Delta \\
\Delta^{*}& \Delta^{*}  & \times \\
\end{array}
\right)$ & $\left(
\begin{array}{ccc}
\times& \Delta & \Delta \\
\Delta^{*} & 0 & \Delta\\
\Delta^{*}& \Delta^{*} & \times \\
\end{array}
\right)$ & $\left(
\begin{array}{ccc}
\times& \Delta & \Delta \\
\Delta^{*} & \times & \Delta\\
\Delta^{*}& \Delta^{*} & 0 \\
\end{array}
\right)$  \\
\hline $P_4$ & $P_5$ & $P_6$  \\
\hline  $\left(
\begin{array}{ccc}
\times& 0 & \Delta \\
0 & \times & \Delta\\
\Delta^{*}& \Delta^{*} & \times \\
\end{array}
\right)$&$\left(
\begin{array}{ccc}
\times& \Delta & 0 \\
\Delta^{*} & \times & \Delta\\
0& \Delta^{*} & \times \\
\end{array}
\right)$ &  $\left(
\begin{array}{ccc}
\times& \Delta & \Delta \\
\Delta^{*} & \times & 0\\
\Delta^{*}& 0 & \times \\
\end{array}
\right)$ \\ \hline
\end{tabular}
\caption{\label{tab1}Possible structures of neutrino mass matrices having texture one zero. where `$\times$ ' stands for non-zero element and real matrix element and each '$\Delta$' for non-zero and complex entity.}
\end{center}
\end{table}\\
Textures $P_{2}$ and $P_{4}$ are related through permutation symmetry to $P_{3}$ and $P_{5}$, respectively \cite{14}. This corresponds to permutation of the 2-3 rows and 2-3 columns of $M_{\nu}$. The corresponding permutation matrix is
\begin{center}
\begin{equation}\label{eq1}
 P_{23} = \left(
\begin{array}{ccc}
    1& 0& 0 \\
  0 & 0 & 1\\
  0& 1& 0 \\
\end{array}
\right).
\end{equation}
\end{center}
As a result of permutation symmetry between different classes, one obtains the following relations among the oscillation parameters
\begin{center}
\begin{equation}\label{eq2}
\theta_{12}^{X}=\theta_{12}^{Y}, \ \
\theta_{23}^{X}=90^{\circ}-\theta_{23}^{Y},\ \
\theta_{13}^{X}=\theta_{13}^{Y}, \ \ \delta^{X}=\delta^{Y} -  180^{\circ},
\end{equation}
\end{center}
where X and Y denote the textures related by 2-3 permutation.\\

In the present work, we attempt to investigate the phenomenological implications of texture
one-zero neutrino mass matrices in the scenario of Det $M_{\nu}$ = 0 or Tr $M_{\nu}$ = 0 condition,
assuming the Dirac nature of neutrinos. Earlier in \cite{29}, we have studied the implication
of Det $M_{\nu}$ = 0 on texture one zero mass matrices for Majorana neutrinos, and found that
normal mass ordering is ruled out for all the six cases of texture one zero mass matrices ,
while only four cases $P_{2}, P_{3}, P_{4}$ and $P_{5}$ are found to be viable for inverted mass ordering at
3$\sigma$ CL. However, in the present work, we find that only two cases $P_2$ and $P_3$ are able to
survive the data for inverted mass ordering, while normal mass ordering remains ruled out
for all the six cases at 3$\sigma$ CL.

The rest of the paper is planned as follows: In section 2, we discuss the methodology used
to reconstruct the neutrino mass matrix for Dirac neutrinos and hence obtain some useful
phenomenological constriants on neutrino masses by imposing texture one zero and zero
determinant (or trace) condition. In section 3, we present the numerical analysis using some
analytical relations. In section 4, we summarize and concludes our work.

\section{Methodology}
In the flavor basis, where charged lepton mass matrix is assumed
to be diagonal, the Dirac neutrino mass matrix $M_{\nu}$,
depending on three neutrino masses ($m_{1}$, $m_{2}$, $m_{3}$) and
the flavor mixing matrix $U$ is expressed as
\begin{equation}\label{eq3}
M_{\nu}=U\left(
\begin{array}{ccc}
\lambda_{1}& 0& 0 \\
0 & \lambda_{2} & 0\\
0& 0& \lambda_{3} \\
\end{array}
\right)U^{\dag},
\end{equation}
where $\lambda_{1} = \eta.m_{1},\lambda_{2} = \kappa.m_{2} ,\lambda_{3} =
m_{3}$ with $\eta.\kappa=\pm1$. The three eigen values ($\lambda_{1}$, $\lambda_{2}$, $\lambda_{3}$)
of a general $3\times3$ hermitian  matrix are real, but not necessarily positive.\\
For the present analysis, we adopt the following parameterization \cite{11}
\begin{equation}\label{eq4}
U=\left(
\begin{array}{ccc}
c_{12}c_{13}& s_{12}c_{13}& s_{13} \\
-c_{12}s_{23}s_{13}-s_{12}c_{23}e^{-i\delta} & -s_{12}s_{23}s_{13}+c_{12}c_{23}e^{-i\delta} & s_{23}c_{13}\\
-c_{12}c_{23}s_{13}+s_{12}s_{23}e^{-i\delta}& -s_{12}c_{23}s_{13}-c_{12}s_{23}e^{-i\delta}& c_{23}c_{13} \\
\end{array}
\right),
\end{equation}
where $c_{ij}$= cos $\theta_{ij}$,  $s_{ij}$= sin $\theta_{ij}$ $(i,j=1, 2, 3)$. \\
If one of the elements of $M_{\nu}$ is considered zero, i.e. $M_{lm} = 0$, it leads to following constraint equation
\begin{equation}\label{eq5}
\eta. m_{1}U_{l1}U_{m1}^{\ast}+\kappa. m_{2}U_{l2}U_{m2}^{\ast}+
m_{3}U_{l3}U_{m3}^{\ast}=0,
\end{equation}
where $l, m$ run over $e, \mu $ and $ \tau$. The solar and atmospheric
mass squared differences ($\delta m^{2}$, $\Delta m^{2}$), where
$\delta m^{2}$ corresponds to solar mass squared difference and
$\Delta m^{2}$ corresponds to atmospheric mass squared
difference, can be defined as
\begin{equation}\label{eq6}
\delta m^{2}=(m_{2}^{2}-m_{1}^{2}),\;
\end{equation}
\begin{equation}\label{eq7}
\Delta m^{2}=|m_{3}^{2}-m_{2}^{2}|,
\end{equation}
then the ratio of two mass-squared differences is given by
\begin{equation}\label{eq8}
R_{\nu}=\dfrac{\delta m^{2}} {|\Delta m^{2}|}.
\end{equation}
The Jarlskog rephrasing parameter $J_{CP}$, which measures the CP violation, is defined as
\begin{equation}\label{eq9}
Im[K_{ij}^{lm}]=J_{CP}\sum_{n}\epsilon_{lmn}
\sum_{k}\epsilon_{ijk},
\end{equation}
where $K_{ij}^{lm}=U_{li}U_{lj}^{\ast}U_{mi}^{\ast}U_{mj}$. The $\varepsilon_{lmn}$ and $\varepsilon_{ijk}$ denote the Levi-Civita symbols. 

Noting that Det $M_{\nu}$=0 if and only if Det $M^{diag}$=0, where $M^{diag}=(\lambda_{1}, \lambda_{2}, \lambda_{3})$, therefore Det $M_{\nu}$=0 implies that one of the eigen values has to be zero. For the normal mass ordering (NO), $m_{1}=0$ and for inverted ordering (IO), $m_{3}$=0. The Tr $M_{\nu}$=0 condition implies  $\eta. m_{1}+\kappa. m_{2}+m_{3}=0$. In the following subsections, we shall study the implication of these conditions on one zero texture, separately.

\subsection{ $M_{lm}=0$ with Det $M_{\nu}$=0}
First of all, we discuss the case of normal mass ordering (NO), which implies $m_{1}=0$. From Eq. (\ref{eq5}), one can obtain the following constraint equation and hence deduce the neutrino mass ratio term $\dfrac{m_{2}}{m_3}$ as
\begin{equation}\label{eq10}
\kappa. m_{2}U_{l2}U_{m2}^{\ast}+
m_{3}U_{l3}U_{m3}^{\ast}=0,
\end{equation}
and
\begin{equation}\label{eq11}
\dfrac{m_{2}}{m_{3}}=-\dfrac{1} {\kappa}\; \dfrac{U_{l3}
U_{m3}^{\ast}}{U_{l2} U_{m2}^{\ast}}.
\end{equation}
In case  $l=m$  (e.g.  the one-zero textures  $P_{1, 2, 3}$),  Eq. (\ref{eq10}) leads to one constraint condition, but we obtain two constraint conditions for $l\neq m$ case (e.g. the one-zero textures  $P_{4, 5, 6}$). In the former case, $\kappa=-1$ must hold since neutrino mass ratios are by definition real and non-negative and
\begin{equation}\label{eq12}
\dfrac{m_{2}}{m_{3}}=-\dfrac{1} {\kappa}\;
\dfrac{|U_{l3}|^{2}}{|U_{l2}|^{2}}.
\end{equation}
In the latter case, we can get two constraint conditions by equating the real and imaginary parts of Eq. (\ref{eq10}) to zero
\begin{equation}\label{eq13}
Re[K_{32}^{lm}]=-\kappa \bigg(\dfrac{m_{2}}{m_{3}}\bigg)\;
|U_{l2}|^{2}|U_{m2}|^{2},
\end{equation}
and
\begin{equation}\label{eq14}
-\kappa \dfrac{1}{\bigg(\dfrac{m_{2}}{m_{3}}\bigg)}\;
\dfrac{1}{|U_{l2}|^{2}|U_{m2}|^{2}}Im[K_{32}^{lm}]=0,
\end{equation}
where $K_{ij}^{lm}=U_{li}U_{lj}^{\ast}U_{mi}^{\ast}U_{mj}$. Using Eqs. (\ref{eq6}) and (\ref{eq7}), neutrino masses ($m_{1}$, $m_{2}$, $m_{3}$) can be expressed in terms of experimentally known mass squared differences ($\delta m^{2}$, $\Delta m^{2}$) as
\begin{equation}\label{eq15}
m_{1}=0, \; m_{2}=\sqrt{\delta m^{2}}, \; m_{3}=\sqrt{\delta
m^{2}+\Delta m^{2}}.
\end{equation}
Hence, we obtain
\begin{equation}\label{eq16}
      \frac{m_{2}} {m_{3}}=\sqrt\frac{R_{\nu}}{1+R_{\nu}}.
\end{equation}
Using Eqs. (\ref{eq11}) and (\ref{eq16}), we can express $R_{\nu}$ in terms of
mixing angles ($\theta_{12}$, $\theta_{23}$, $\theta_{13}$) and
Dirac CP-violating phase ($\delta$) as
\begin{equation}\label{eq17}
R_{\nu}=\bigg[\bigg(\dfrac{U_{l2}U_{m2}^{\ast}} {U_{l3}
U_{m3}^{\ast}}\bigg)^{2}-1\bigg]^{-1}.
\end{equation}

In case of inverted mass ordering (IO), which implies $m_{3}$=0, one  obtain the following constraint equation using Eq. (\ref{eq5}),and hence deduce the neutrino mass ratio term $\dfrac{m_{2}}{m_{1}}$ as
\begin{equation}\label{eq18}
\eta. m_{1}U_{l1}U_{m1}^{\ast}+\kappa.
m_{2}U_{l2}U_{m2}^{\ast}=0,
\end{equation}
\begin{equation}\label{eq19}
\dfrac{m_{2}}{m_{1}}=-\dfrac{\eta}{\kappa}\; \dfrac{U_{l1}
U_{m1}^{\ast}}{U_{l2} U_{m2}^{\ast}}.
\end{equation}\\
Since mass ratio term $\dfrac{m_{2}}{m_{1}}$  is by definition real and non-negative, therefore $\eta=\pm1, \kappa=\mp1$ must hold. Using Eq. (\ref{eq19}), one can deduce a constraint equation in case of $l=m$ (e.g. the one-zero textures  $P_{1, 2,3} $) in terms of mass ratio $\dfrac{m_{2}}{m_{1}}$
\begin{equation}\label{eq20}
\dfrac{m_{2}}{m_{1}}=-\dfrac{\eta}{\kappa}\;
\dfrac{|U_{l1}|^{2}}{|U_{l2}|^{2}}.
\end{equation}
For  $l\neq m $, one can equate the real and imaginary parts of Eq. (\ref{eq18}) to zero and hence obtain the two constraint equations
\begin{equation}\label{eq21}
Re[K_{12}^{lm}]=-\dfrac{\eta}{\kappa} \bigg(\dfrac{m_{2}}{m_{1}}\bigg)\;
|U_{l2}|^{2}|U_{m2}|^{2},
\end{equation}
\begin{equation}\label{eq22}
-\dfrac{\eta}{\kappa} \dfrac{1}{\bigg(\dfrac{m_{2}}{m_{1}}\bigg)}\;
\dfrac{1}{|U_{l2}|^{2}|U_{m2}|^{2}}Im[K_{12}^{lm}]=0.
\end{equation}
The neutrino mass spectrum is given as
\begin{equation}\label{eq23}
m_{1}=\sqrt{\Delta m^{2}-\delta m^{2}},\; m_{2}=\sqrt{\Delta
m^{2}},\; m_{3}=0, \;
\end{equation}
The non-zero and finite mass ratio $\dfrac{m_{2}}{m_{1}}$  can be
related to $R_{\nu}$ as
\begin{equation}\label{eq24}
      \dfrac{m_{2}} {m_{1}}=\dfrac{1} {\sqrt{1-R_{\nu}}}.
\end{equation}
Using Eqs. (\ref{eq19}) and (\ref{eq24}), we can express $R_{\nu}$ in terms of
mixing angles ($\theta_{12}$, $\theta_{23}$, $\theta_{13}$) and
Dirac CP violating phase ($\delta$) as
\begin{equation}\label{eq25}
R_{\nu}=1- \bigg(\dfrac{U_{l2}U_{m2}^{\ast}} {U_{l1}
U_{m1}^{\ast}}\bigg)^{2}.
\end{equation}
The Jarlskog rephrasing invariant parameter
$J_{CP}$, which measures the CP violation, is defined as
\begin{equation}\label{eq26}
Im[K_{ij}^{lm}]=J_{CP}\sum_{n}\epsilon_{lmn}
\sum_{k}\epsilon_{ijk}.
\end{equation}
In case of $M_{lm}$=0   with $m_{1}= 0$, where $l\neq m$,
Eq. (\ref{eq14}) leads to either $\dfrac{m_{3}}{m_{2}}=0$ or
$\dfrac{1}{|U_{l2}|^{2}|U_{m2}|^{2}}=0$ or $Im[K_{32}^{lm}]=0$.
From these possibilities, $\dfrac{m_{3}}{m_{2}}=0$ implies $m_{3}=0$. With the help of Eq. (\ref{eq5}), we obtain, $m_{1}= m_{2}= m_{3}=0$, which is in contradiction with the solar neutrino oscillation data (i.e. $m_{2}>m_{1}$ ) \cite{35,36}. Moreover, the elements of mixing matrix $U$ are always non-zero and finite,
so we are left with $Im[K_{32}^{lm}]=0$, which implies
$J_{CP}=0$. Therefore, CP violation is only possible for the
textures $P_{1, 2, 3}$ with $m_{1}= 0$, while $\delta=0^{0}$ or
$180^{0}$ holds for remaining one-zero textures viz. $P_{4, 5, 6}$.
Similarly, in case of $M_{lm}$=0  with $m_{3}= 0$, where $l\neq m$,
Eq. (\ref{eq22}) leads to either $\dfrac{m_{1}}{m_{2}}=0$ or $\dfrac{1}{|U_{l2}|^{2}|U_{m2}|^{2}}=0$ or $Im[K_{12}^{lm}]=0$. Here $\dfrac{m_{1}}{m_{2}}=0$ implies $m_{1}= 0$. Using Eq. (\ref{eq5}), we find $m_{1}= m_{2}= m_{3}=0$, which again contradicts the inequality relation $\delta m^{2}>0$ or equivalently $m_{2}>m_{1}$ as established by the solar neutrino experiments. Therefore, we have only $Im[K_{12}^{lm}]=0$ which implies $J_{CP}=0$. Hence CP violation
holds only for textures $P_{1, 2, 3}$ with $m_{3}=0$, while $\delta=0^{0}$ or
$180^{0}$ holds for remaining one-zero textures viz.  $P_{4, 5, 6}$.

\subsection{$M_{lm}=0$ with Tr $M_{\nu}$=0}
The second basis independent condition is Tr $M_{\nu}$=0. The zero trace implies the sum of three neutrino eigen values of $M_{\nu}$ must be zero
\begin{equation}\label{eq27}
\eta. m_{1}+\kappa. m_{2}+m_{3}=0
\end{equation}
Using Eqs. (\ref{eq5}) and (\ref{eq27}), we obtain the following relations for neutrino mass ratios
\begin{equation}\label{eq28}
\alpha\equiv\dfrac{m_{1}}{m_{3}}=\dfrac{1}{\eta}\;\dfrac{U_{l2}U_{m2}^{\ast}-U_{l3}U_{m3}^{\ast}}{U_{l1}U_{m1}^{\ast}-U_{l2}U_{m2}^{\ast}},
\end{equation}
\begin{equation}\label{eq29}
\beta\equiv\dfrac{m_{2}}{m_{3}}=\dfrac{1}{\kappa}\;\dfrac{U_{l3}U_{m3}^{\ast}-U_{l1}U_{m1}^{\ast}}{U_{l1}U_{m1}^{\ast}-U_{l2}U_{m2}^{\ast}}.
\end{equation}
Since both $\alpha$ and $\beta$ are by definition real and non-negative, the imaginary parts of the quantities
on the right-hand side of Eqs. (\ref{eq28}) and (\ref{eq29}) have to disappear. This requirement may lead us to the
determination of the CP violating phase $\delta$, as we shall show below.\\
For $M_{lm}=0$, where $l\neq m$ (e.g. textures  $P_{4, 5, 6}$ ),  one can  again show  that CP violation is forbidden.  Using Eqs. (\ref{eq5}) and (\ref{eq27}) again  and subsequently  equating the imaginary part to zero, we obtain the  following  constraint equation
\begin{equation}\label{eq30}
(2\eta \alpha+\kappa \beta)Im(K_{12}^{lm})=0.
\end{equation}
Eq. (\ref{eq30}) implies either $(2\eta \alpha+\kappa \beta)=0$   or $Im(K_{12}^{lm})=0$. On solving $(2\eta \alpha+\kappa \beta)=0$ and  Eq. (\ref{eq27}) simultaneously, we obtain $m_{3}=\eta.\; m_{1}$, which is not possible. Therefore, we are left with $Im(K_{12}^{lm})=0$,   which implies $J_{CP}=0$. Hence CP violation is only possible for the textures $P_{1, 2, 3}$  and $\delta=0^{0}$ or $180^{0}$ for rest of the one zero textures. Therefore, it is concluded here that in both the scenarios, namely Det $M_{\nu}$=0 and Tr $M_{\nu}$=0, textures with vanishing diagonal entry lead to CP violation, while textures with vanishing off diagonal entry lead to CP conservation. \\
The ratio of two mass-squared differences $R_{\nu}$ and neutrino mass spectrum ($m_{1}, m_{2}, m_{3}$)  in terms of neutrino mass ratios $\alpha $ and $ \beta $  can be given as
\begin{equation}\label{eq31}
R_{\nu}=\dfrac{\delta m^{2}} {|\Delta m^{2}|} =\dfrac{(\beta ^{2}-\alpha ^{2})}{\left |1-\beta^{2}  \right |},
\end{equation}
\begin{equation}\label{eq32}
m_{3}=\sqrt{\dfrac{\delta m^{2}}{\beta^{2}-\alpha^{2}}},  \qquad m_{2}=m_{3} \beta, \qquad m_{1}=m_{3} \alpha.
\end{equation}
It must be noted that Eqs.(\ref{eq17}), (\ref{eq25}), (\ref{eq31})  provide a very useful
constraint to restrict the parameter space of neutrino
oscillation parameters.\\
\begin{table}
\begin{small}
\begin{center}
\begin{tabular}{|c|c|c|}
  \hline
  Parameter& Best Fit & 3$\sigma$ \\
  \hline
   $\delta m^{2}$ $[10^{-5}eV^{2}]$ & $7.50$& $7.03$ - $8.09$  \\
   \hline
   $|\Delta m^{2}_{31}|$ $[10^{-3}eV^{2}]$ (NO) & $2.52$ & $2.407$ - $2.643$ \\
   \hline
  $|\Delta m^{2}_{31}|$ $[10^{-3}eV^{2}]$ (IO) & $2.52$ &  $2.39$ - $2.63$ \\
  \hline
  $\theta_{12}$ & $33.56^{\circ}$ & $31.3^{\circ}$ - $35.99^{\circ}$\\
  \hline
  $ \theta_{23}$ (NO) & $41.6^{\circ}$  & $38.4^{\circ}$ - $52.8^{\circ}$ \\
  \hline
  $\theta_{23}$ (IO)& $50.0^{\circ}$ &  $38.8^{\circ}$ - $53.1^{\circ}$ \\
  \hline
  $\theta_{13}$ (NO) & $8.46^{\circ}$ &  $7.99^{\circ}$ - $8.90^{\circ}$ \\
  \hline
  $\theta_{13}$ (IO) & $8.49^{\circ}$ &  $8.03^{\circ}$ - $8.93^{\circ}$ \\
  \hline
  $\delta$ (NO) & $261^{\circ}$ & $0^{\circ}$ - $360^{\circ}$ \\
  \hline
  $\delta$ (IO) &$277^{\circ}$& $145^{\circ}$ - $391^{\circ}$ \\
\hline
\end{tabular}
\caption{\label{tab2}Current neutrino oscillation parameters from global fits at 3$\sigma$ confidence level(CL)  \cite{37}. NO (IO) refers to normal (inverted) neutrino mass ordering.}
\end{center}
\end{small}
\end{table}

\section{Numerical analysis}
For the purpose of numerical calculations, we have used the 3$\sigma$ values of the lepton mixing angles as well as neutrino mass square differences as listed in Table \ref{tab2}. To start with, we span the parameter space of input neutrino oscillation parameters ($\theta_{12},\theta_{23}, \theta_{13}, \delta m^{2}, \Delta m^{2}$) by choosing the randomly generated points of the order of $10^{7}$. Assuming the Dirac nature of neutrinos, we classify the six possible one-zero textures into two categories viz. CP violating textures ($P_{1}, P_{2}, P_{3}$) and CP conserving textures ($P_{4}, P_{5}, P_{6}$), while for CP violating textures ($P_{1}, P_{2}, P_{3}$), Dirac CP violating phase ($\delta$) is allowed to vary between [$0^{0}, 360^{0}$] at 3$\sigma$ CL.  For CP conserving textures ($P_{4}, P_{5}, P_{6}$), only $\delta =0^{0}$ or $180^{0}$ are allowed. Using Eqs. (\ref{eq17}), (\ref{eq25}), (\ref{eq31}), the parameter space of Dirac CP violating phase ($\delta$) can be subsequently constrained.  The present numerical analysis is divided into two parts : Firstly, we investigate the phenomenological consequences of zero determinant condition on one zero textures. The zero determinant condition implies either $m_{1}=0$ or $m_{3}=0$, corresponding to normal and inverted mass ordering, respectively. As a next step, we study the implication of zero trace for the same.  In order to add more understanding to the phenomenological results, the approximate relation of mass ratios and $R_{\nu}$ have been taken into account up to the leading order term of sin$\theta_{13}$. We emphasize here that the present numerical analysis is based on the exact formula not on approximations The exact analytical relations for neutrino mass ratios for texture one zero mass matrices along with Det$M_{\nu}$=0 or Tr $M_{\nu}$=0 have been summarized in Table \ref{tab3} \ref{tab4} \ref{tab5} .

\begin{center}
\large{(A)\; \;\textbf{ $M_{lm}=0$ with Det $M_{\nu}=0$}}
\end{center}
\subsection{ CP violating textures ($P_{1}$, $P_{2}$, $P_{3}$)}

\subsubsection{Texture $P_{1}$ with vanishing $m_{1}$ and $m_{3}$}

For texture $P_{1}$ with $m_{1}=0$, one can obtain the full
analytical expressions for mass ratio
($\dfrac{m_{2}}{m_{3}}$ ) and $R_{\nu}$ term from Eqs. (\ref{eq11}) and (\ref{eq17})
\begin{equation}\label{eq33}
\dfrac{m_{2}}{m_{3}}= -\dfrac{1}{\kappa}\dfrac{t_{13}^{2}}{s_{12}^{2}},
\end{equation}
\begin{equation}\label{eq34}
R_{\nu}=\dfrac{{t}_{13}^{4}}{{s}_{12}^{4}-{t}_{13}^{4}},
\end{equation}
where $\kappa=-1$ holds to ensure that $\dfrac{m_{2}}{m_{3}}$ is non-negative.
Taking into account the $3\sigma$ experimental range of
($\theta_{12}$, $\theta_{13}$), we find that $R_{\nu}$
excludes the current experimental range. Similarly for
$m_{3}$=0, we obtain  $R_{\nu}=1-t_{12}^{4}=0.63-0.85$ from Eq. (\ref{eq25}), which is again
inconsistent with experimental range of $R_{\nu}$ . Hence, texture
$P_{1}$ is ruled out with current experimental data for both
$m_{1}=0$ and $m_{3}=0$ cases.

\subsubsection{Texture $P_{2}$ with vanishing $m_{1}$ and $m_{3}$}
\begin{figure}[h!]
\begin{center}
\subfigure[]{\includegraphics[width=0.4\columnwidth]{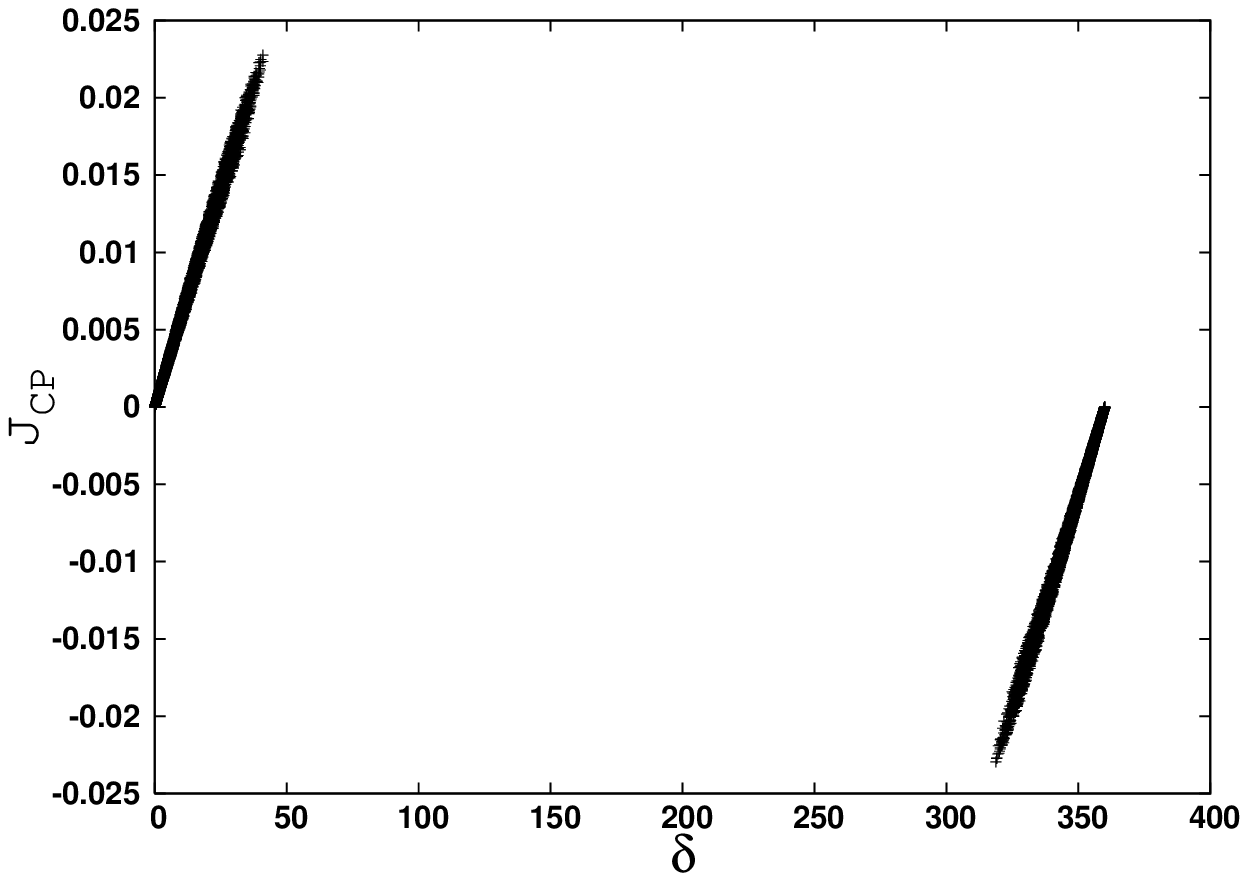}}
\subfigure[]{\includegraphics[width=0.4\columnwidth]{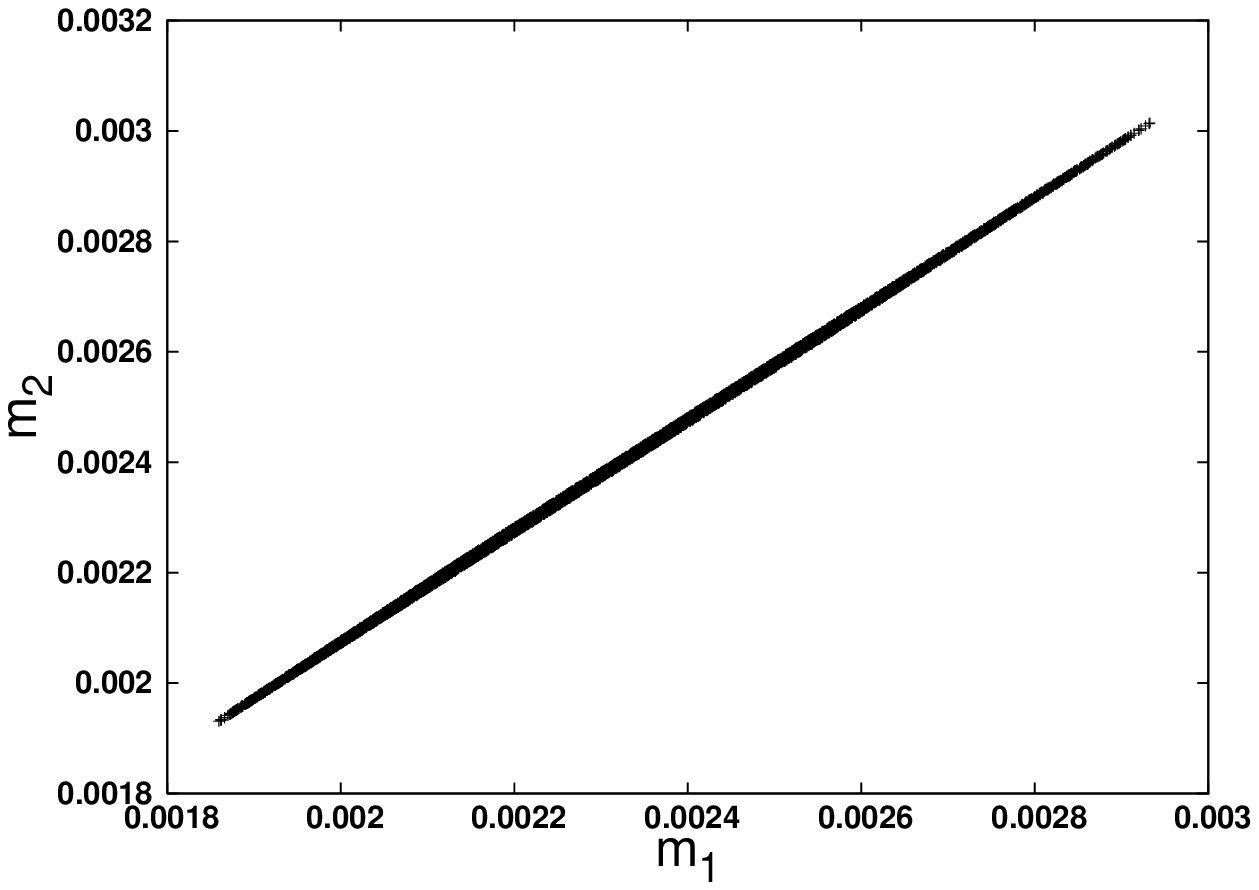}} \ \ \
\caption{\label{fig1} Correlation plots for texture $P_{2}$ with $m_{3}=0$. The masses $m_{1}$ and $m_{2}$ are measured in eV.}
\end{center}
\end{figure}

For texture $P_{2}$ with $m_{1}=0$, we obtain the following
relations in the leading order term of $\theta_{13}$
\begin{equation}\label{eq35}
\dfrac{m_{2}}{m_{3}}\approx -\dfrac{1}{\kappa}.\dfrac
{t_{23}^{2}}{c_{12}^{2}},
\end{equation}

\begin{equation}\label{eq36}
R_{\nu}\approx
\dfrac{t_{23}^{4}}{c_{12}^{4}-t_{23}^{4}},
\end{equation}
where $\kappa=-1$ holds so as to get non-negative $\dfrac{m_{2}}{m_{3}}$. Using
$3\sigma$ experimental range of mixing angles ($\theta_{12}$,
$\theta_{23}$, $\theta_{13}$), $R_{\nu}$ turns out to be well above the current experimental range. Therefore, texture $P_{2}$ with $m_{1}=0$ is not
consistent with the neutrino oscillation data at $3\sigma$ CL. On the contrary
, texture $P_{2}$ with $m_{3}=0$ is found to be
consistent with current experimental data at $3\sigma$ CL. The
analytical expressions for mass ratios $\dfrac{m_{2}}{m_{1}}$  and
$R_{\nu}$ (up to the leading $s_{13}$ term) are presented below
\begin{equation}\label{eq37}
\dfrac{m_{2}}{m_{1}}\approx -\dfrac{\eta}{\kappa}. t_{12}^{2}\bigg(1+\dfrac{2
c_{\delta}s_{13}t_{23}}{s_{12}c_{12}}\bigg),
\end{equation}
\begin{equation}\label{eq38}
R_{\nu}\approx  1-\dfrac{1}{t_{12}^{4}}\bigg(1-\dfrac{4
c_{\delta}s_{13}t_{23}}{s_{12}c_{12}}\bigg),
\end{equation}
where $c_{\delta}$ $\equiv$ $cos\delta$. Here, $\eta=\pm1, \kappa=\mp1$ must hold so as to get non-negative $\dfrac{m_{2}}{m_{1}}$. Since $\delta m^{2}>0$ or equivalently $m_{2} > m_{1}$, we have cos$\delta>0$ from Eq. (\ref{eq37}), which implies that
Dirac CP violating phase ($\delta$) lies in the first and fourth
quadrant i.e. $\delta< 90^{0}$ and $\delta>270^{0}$. From figure \ref{fig1}(a), it is evident that parameter space of Dirac CP phase lies in
the range,  $\delta \in [0^{0},56^{0}]\oplus[306^{0},360^{0}]$. The correlation plot for ($m_{1}, m_{2}$) indicates that there is strong linear correlation between $ m_{1}$ and $ m_{2}$ [figure \ref{fig1}(b)].  The
parameter space of ($J_{CP}, \delta$) shows that $J_{CP}\neq0$,
indicating the CP violation [figure \ref{fig1}(a)].
\subsubsection{Texture $P_{3}$ with vanishing $m_{1}$ and $m_{3}$ }
\begin{figure}[h!]
\begin{center}
\subfigure[]{\includegraphics[width=0.4\columnwidth]{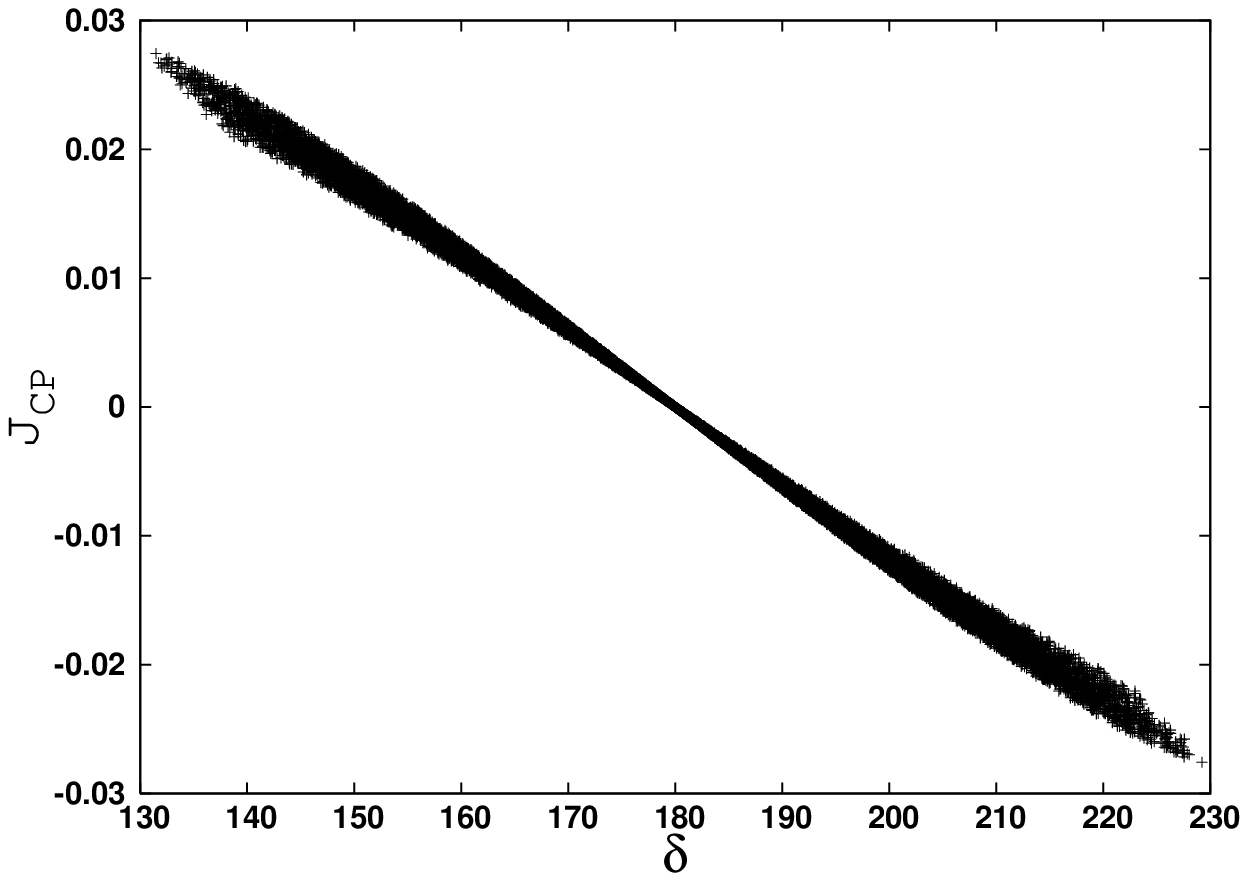}}
\subfigure[]{\includegraphics[width=0.4\columnwidth]{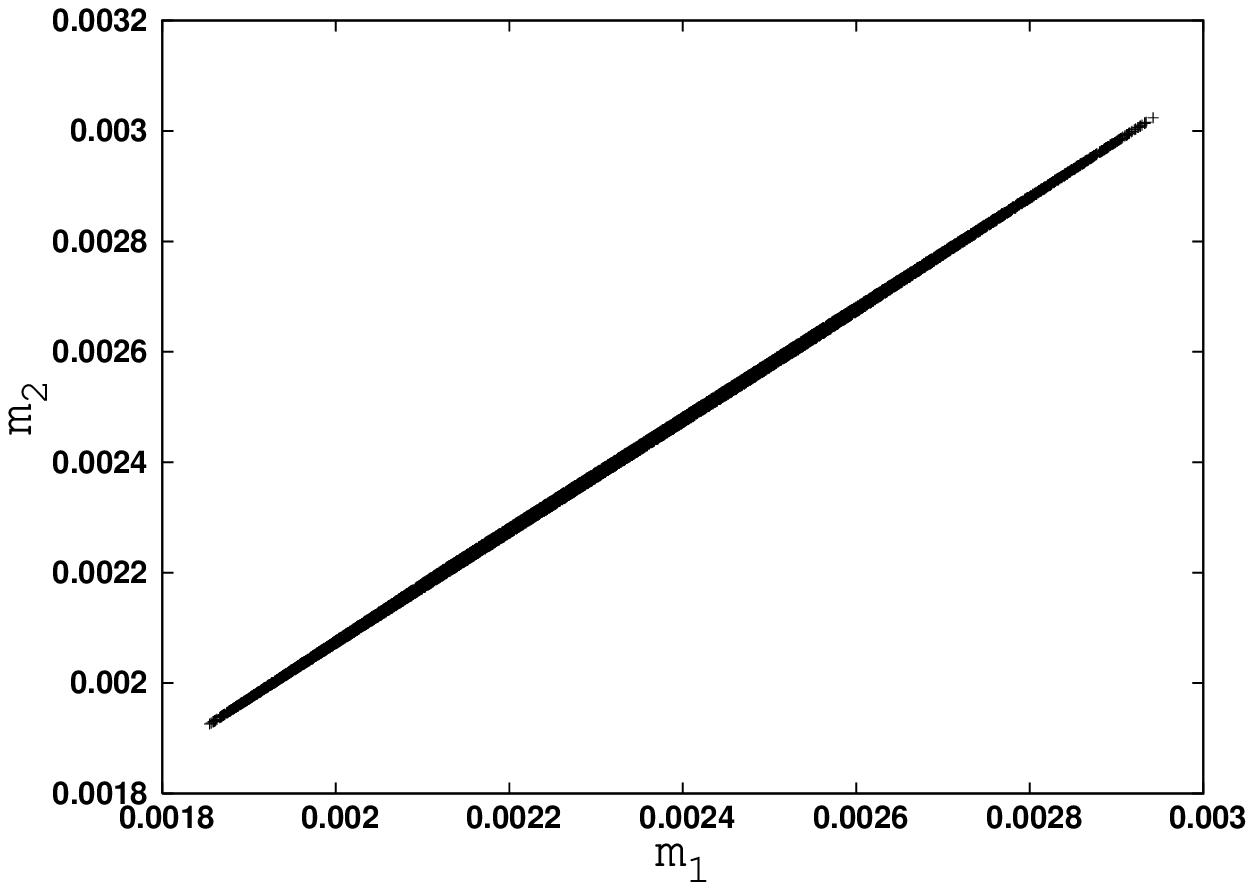}}\\
\caption{\label{fig2}Correlation plots for textures $P_{3}$ with $m_{3}=0$. The masses $m_{1}$ and $m_{2}$ are measured in eV. }
\end{center}
\end{figure}

As already pointed out in Section 2, the textures $P_{3}$ and  $P_{2}$ are related due to 2-3 permutation symmetry.
Therefore, texture $P_{3}$ with $m_{1}=0$ also remains inconsistent with current experimental data.
Similar to the previous case for texture $P_{2}$, texture $P_{3}$ with $m_{3}=0$ also favors the current neutrino oscillation data.
With the help of Eqs. (\ref{eq19}) and (\ref{eq25}), we deduce some analytical expressions in the leading order of $s_{13}$ term.
\begin{equation}\label{eq39}
\dfrac{m_{2}}{m_{1}}\approx -\dfrac{\eta}{\kappa} t_{12}^{2}\bigg(1-\dfrac{2
c_{\delta}s_{13}}{s_{12}c_{12}t_{23}}\bigg),
\end{equation}
\begin{equation}\label{eq40}
R_{\nu}\approx 1-\dfrac{1}{t_{12}^{4}}\bigg(1+\dfrac{4
c_{\delta}s_{13}}{s_{12}c_{12}t_{23}}\bigg).
\end{equation}
where $\eta=\pm1, \kappa=\mp1$. From Eq. (\ref{eq39}), we have cos $\delta<0$ in view of the
fact that $m_{2} > m_{1}$, which implies that
Dirac CP violating phase ($\delta$) lies in the second and third
quadrant i.e. $90^{0}<\delta<270^{0}$. From figure \ref{fig2}(a), it is
evident that parameter space of Dirac CP phase lies in the range, $\delta \in
[130^{0},230^{0}]$. The allowed parameter space of $\delta$ can be further verified by using the relation,  $\delta$ (for  texture $P_{3}$)= $\delta$ (for  texture $P_{2}$)$\pm 180^{0}$, resulting from the permutation symmetry. The correlation plot between $m_{1}$ and $m_{2}$ exhibits a linear correlation [figure \ref{fig2}(b)]. The CP-violation (implying $J_{CP} \neq 0$) in texture $P_{3}$ can be seen  in figure \ref{fig2}(a) along with vanishing value of $J_{CP}$.\\

\begin{figure}[h!]
\begin{center}

\subfigure[]{\includegraphics[width=0.4\columnwidth]{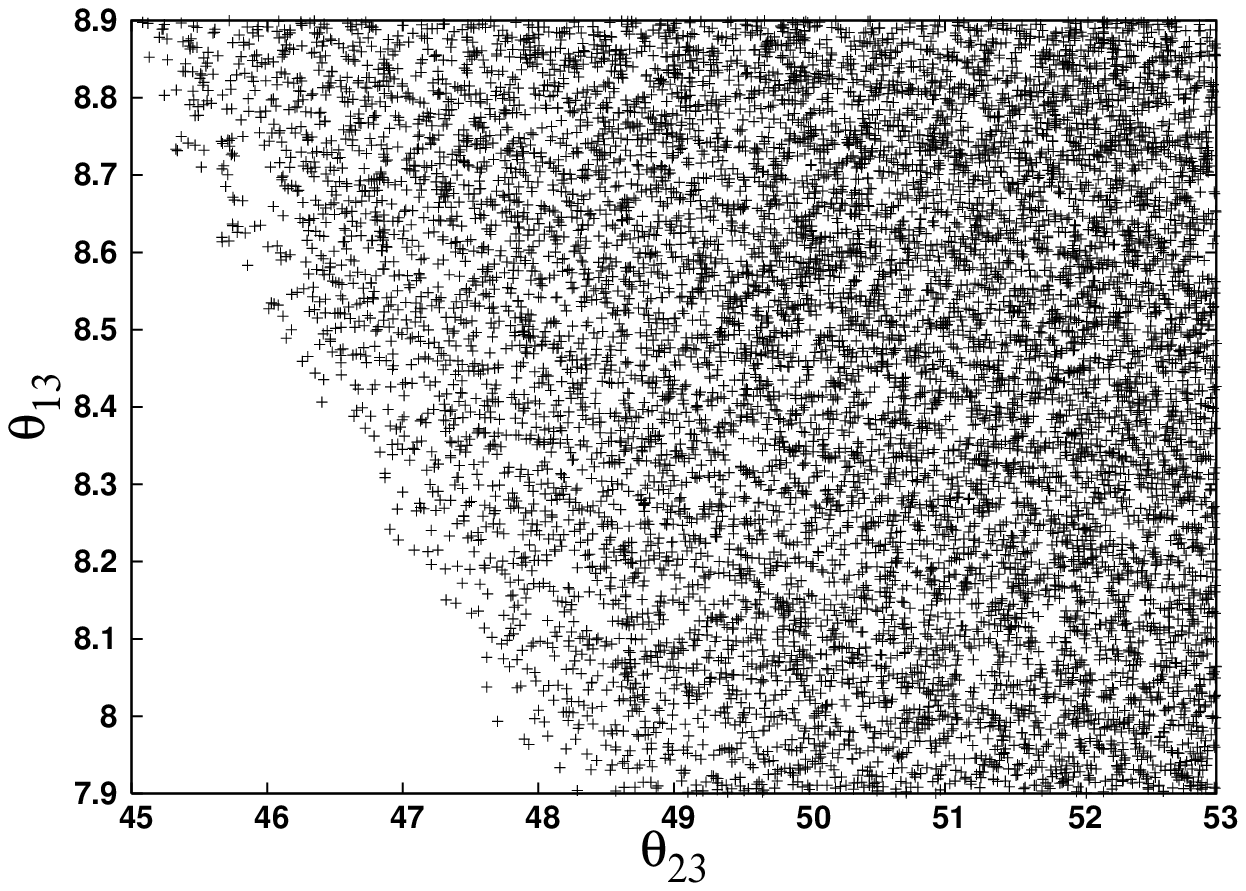}}
\subfigure[]{\includegraphics[width=0.4\columnwidth]{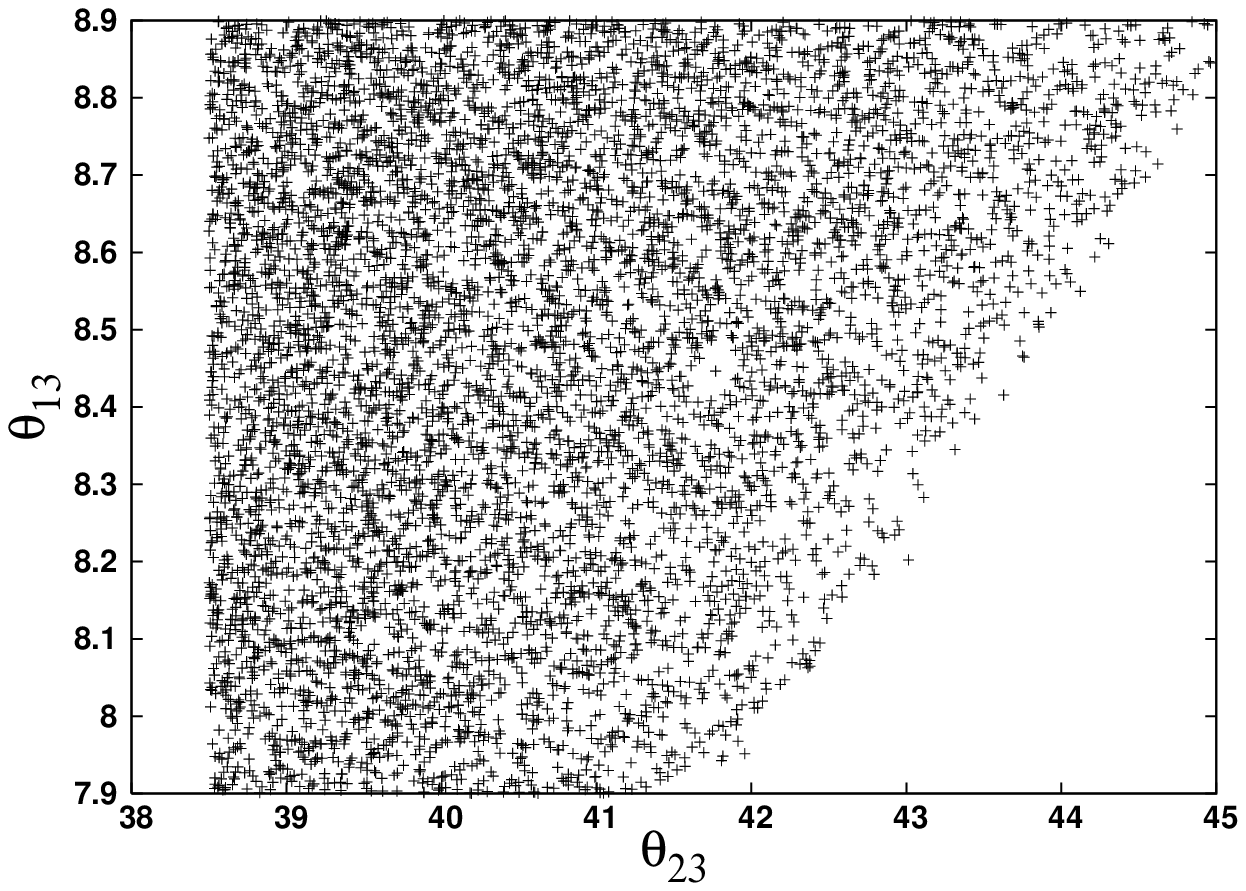}}

\caption{\label{fig3} Correlation plots for textures (a) $P_{2}$ with $m_{3}=0$  and (b) $P_{3}$  with $m_{3}=0$,  depicting the permutation symmetry. }
\end{center}
\end{figure}

 In figures \ref{fig3}(a) and \ref{fig3}(b), we have explicitly shown the  permutation symmetry between textures $P_2$ and $P_3$. Also, we can see that  texture $P_{2}$ allows only upper octant of $\theta_{23}$ (i.e. $\theta_{23}>45^{0})$, while texture $P_{3}$  allows only lower octant (i.e. $\theta_{23}<45^{0})$.  For higher values of reactor angle $\theta_{13}$, $\theta_{23}$ is found to shift towards $45^{0}$.  Figures \ref{fig3}(a) and \ref{fig3}(b) may appear to show slight deviation from the permutation symmetry relation:
\begin{equation}\label{eq41}
\theta_{23}^{P_3}=90^{\circ}-\theta_{23}^{P_2}.
\end{equation}
However, this apparent deviation is just because the experimentally allowed 3$\sigma$ range for $\theta_{23}$ is not symmetric around $\theta_{23} = 45^\circ$.\\
The NO$\nu$A experiment has recently excluded the maximal-mixing value $\theta_{23} = 45^{0}$ at the 2.6$\sigma$ confidence level \cite{38}, hence hints towards the non-maximality of $\theta_{23}$.  In Ref. \cite{39,40,41} a slight preference for the upper octant (more pronounced in IO) has been indicated, although both octants are allowed at 2$\sigma$ CL. The further robust measurement is needed to decide the octant of $\theta_{23}$ and hence the compatibility of above textures.
 
\subsection{ CP conserving textures ($P_{4}, P_{5}, P_{6}$)}
\subsubsection{Texture $P_{4}$ with vanishing $m_{1}$ and $m_{3}$ }
With the help of Eq. (\ref{eq11}), we obtain
\begin{equation}\label{eq42}
\dfrac{m_{2}}{m_{3}}=-\dfrac{1}{\kappa}\;\dfrac{s_{13}s_{23}} {-s_{12}^{2} s_{23}s_{13}\pm s_{12}c_{12}c_{23}},
\end{equation}
where $\kappa=-1$. The upper and lower signs correspond to $\delta =0^{0}$ and
$180^{0}$, respectively.  Using Eqs. (\ref{eq11}) and (\ref{eq17}), one can deduce
the mass ratio $\dfrac{m_{2}}{m_{3}}$  and $R_{\nu}$ terms
\begin{equation}\label{eq43}
\dfrac{m_{2}}{m_{3}}\approx
\dfrac{t_{23}s_{13}}{s_{12}c_{12}},
\end{equation}
\begin{equation}\label{eq44}
R_{\nu}\approx \dfrac{t_{23}^{2}s_{13}^{2}}{s_{12}^{2}c_{12}^{2}},
\end{equation}

where $s_{13}$ is expanded in leading order approximation. From above equation, it is clear that $R_{\nu}\propto s_{13}^{2}$, depends
strongly on reactor mixing angle $\theta_{13}$. The latest mixing
data (at $3\sigma$ CL) leads to rather large $R_{\nu}$, lying in
the range [0.05, 0.5] and hence excluded by current experimental
range of  $R_{\nu}$.

On the other hand, for $m_{3}=0$, we obtain the following relations (in the leading order of $s_{13}$ term)
from Eqs. (\ref{eq19}) and (\ref{eq25}),
\begin{equation}\label{eq45}
\dfrac{m_{2}}{m_{1}}\approx -\dfrac{\kappa}{\eta}. \bigg(1\pm
\dfrac{t_{23}s_{13}}{s_{12}c_{12}}\bigg),
\end{equation}
where $\eta.\kappa=-1$ must hold. The upper and lower signs refer
to the cases of $\delta=0^{0}$ and $180^{0}$, respectively. From Eq. (\ref{eq45}), $\delta= 180^{0}$ is disallowed since it leads to $m_{2}<m_{1}$. Hence, for $\delta=0^{0}$, we obtain
\begin{equation}\label{eq46}
R_{\nu} \approx \dfrac{2{t}_{23} s_{13}}{{s}_{12}{c}_{12}}.
\end{equation}
Using $3\sigma$  range of mixing angles, we obtain $0.467\leq
R_{\nu}\leq 0.963$, which is again in conflict with the
experimental range of $R_{\nu}$. Therefore,
texture $P_{4}$ is not consistent with the neutrino oscillation data
at $3\sigma$ CL, nor is the texture $P_{5}$ due to permutation
symmetry.

\subsubsection{Texture $P_{6}$ with vanishing $m_{1}$ and $m_{3}$ }

As already discussed in Section 3, there is no CP violation in this case. Therefore, only $\delta=0^{0} $or
$ 180^{0}$ is allowed. For texture $P_{6}$ with $m_{1}=0$, we can obtain the full analytical expression for mass ratio ($\frac{m_{2}}{m_{3}}$) using Eq. (\ref{eq11})
\begin{equation}\label{eq47}
\dfrac{m_{2}}{m_{3}}= -\dfrac{1}{\kappa}.\;\dfrac{s_{23} c_{23} c_{13}^{2}}{(s_{12}^{2}s_{13}^{2}-c_{12}^{2})c_{23}s_{23}+s_{12}c_{12}s_{13}(\pm s_{23}^{2} \mp c_{23}^{2})},
\end{equation}
where $\kappa=-1$ holds so as to get the non-negative mass
ratio term $\dfrac{m_{2}}{m_{3}}$.  The upper and lower signs
correspond to $\delta=0^{0}$ and $180^{0}$, respectively. In the
leading order approximation of $s_{13}$, one can deduce the mass ratio
($\dfrac{m_{2}}{m_{3}}$) as
\begin{equation}\label{eq48}
\dfrac{m_{2}}{m_{3}}\approx \frac{1}{\kappa}.\dfrac
{1}{c_{12}^{2}},
\end{equation}
where $\kappa=1$. From above equation, we find that $m_{2}>m_{3}$, which is not possible in case of normal mass ordering ($m_{1}=0, m_{2}<m_{3}$).  Similarly, it is found that texture
$P_{6}$ with $m_{3}=0$ also remains incompatible with current neutrino
oscillation data \cite{34}. With the help of Eq. (\ref{eq19}), one can derive
the full analytical expression for $\frac{m_{2}}{m_{1}}$
\begin{equation}\label{eq49}
\dfrac{m_{2}}{m_{1}}=-\dfrac{\eta}{\kappa}\;\dfrac{s_{23}c_{23}(c_{12}^{2}s_{13}^{2}-s_{12}^{2})+s_{12}c_{12}s_{13}(\pm c_{23}^{2} \mp s_{23}^{2})}{s_{23}c_{23}(s_{12}^{2}s_{13}^{2}-c_{12}^{2})+s_{12}s_{13}c_{12}(\pm s_{23}^{2} \mp c_{23}^{2})},
\end{equation}
where $\eta.\kappa=-1$. The upper and lower signs in the above
expression refer to the cases  $\delta=0^{0}$ and $180^{0}$,
respectively. In the leading order approximation of $s_{13}$, we
deduce the mass ratio ($\dfrac{m_{2}}{m_{1}}$) and  $R_{\nu}$ as
\begin{equation}\label{eq50}
\dfrac{m_{2}}{m_{1}}\approx t_{12}^{2}\bigg(1\mp\dfrac{2s_{13}}{t_{2(23)}s_{12}c_{12}}\bigg),
\end{equation}
\begin{equation}\label{eq51}
R_{\nu}\approx 1-\dfrac{1}{t_{12}^{4}}\bigg(1\pm \dfrac{4
s_{13}}{t_{2(23)}s_{12}c_{12}}\bigg).
\end{equation}
From the above equation, we find that $m_{2}<m_{1}$ for both $\delta=0^{0}$ and $180^{0}$, which contradicts the solar neutrino oscillation data. Therefore, texture $P_{6}$ with $m_{3}=0$ is ruled out with latest experimental data.

For sake of completion, we have provided the hermitian mass matrices for  two allowed cases $P_{2}$  and $P_{3}$ of texture	one zero with Det $M_{\nu}$=0 condition.\\
\begin{equation*}
M_{\nu}^{P_{2}}=
\begin{pmatrix}
a_{11} & a_{12} & a_{13}\\
a_{21} & a_{22} & a_{23}\\
a_{31} & a_{32} & a_{33}
\end{pmatrix},
\end{equation*}
where
\begin{align*}
&a_{11}=0.0104-0.0211,\\
&a_{12}=((-0.0402)-(-0.0167))+i((-0.0252)-0.0250),\\
&a_{13}=(0.0183-0.0374)+i((-0.0314)-0.0317),\\
&a_{21}=((-0.0402)-(-0.0167))-i((-0.0250)-0.0250), \\
&a_{22}=0.0,\\
&a_{23}=(0.00460-0.00907)+i((-0.00700)-0.00682),\\
&a_{31}=(0.0183-0.0374)-i((-0.0314)-0.0317),\\
&a_{32}=(0.00460-0.00907)-i((-0.00700)-0.00682),\\
&a_{33}=(-0.0219)-(-0.0110),
\end{align*}
and
\begin{equation*}
M_{\nu}^{P_{3}}=
\begin{pmatrix}
b_{11} & b_{12} & b_{13}\\
b_{21} & b_{22} & b_{23}\\
b_{31} & b_{32} & b_{33}
\end{pmatrix},
\end{equation*}
where
\begin{align*}
&b_{11}=0.0102-0.0194,\\
&b_{12}=(0.0199- 0.0364)+i(-0.0288-0.0297),\\
&b_{13}=((-0.0405)- (-0.0192))+i((-0.0247)-0.0236),\\
&b_{21}=(0.0199- 0.0364)-i((-0.0288)-0.0297) , \\
&b_{22}= (-0.0201)- (-0.0111),\\
&b_{23}=(0.00492-0.00907)+i(-0.00640-0.00660),\\
&b_{31}=((-0.0405)-(-0.0192))-i((-0.0246)-0.0236),\\
&b_{32}=(0.00492-0.00907)-i((-0.00640)-0.00660),\\
&b_{33}=0.0.
\end{align*}

\begin{center}
\large{(B)\; \;\textbf{ $M_{lm}=0$ with Tr $M_{\nu}=0$}}
\end{center}
After having discussed the texture one zero with Det $M_{\nu}$=0, we discuss the texture one zero with Tr $M_{\nu}$=0. It is found from our analysis that normal mass ordering is ruled out for all the six  of neutrino mass matrix $M_{\nu}$, while two of them (i.e. $P_{2}$ and $P_{3}$) in the case of inverted mass ordering are found to be compatible with experimental data at 3$\sigma$ level. The survived textures are phenomenologically related to each other due to 2-3 permutation symmetry. From the analysis, we find that in case of $M_{lm}=0$ with Tr $M_{\nu}=0$, where  $l=m$, only $\eta = +1$ and $\kappa = -1$ possibility holds in case of textures $P_{2}$ and $P_{3}$ as given in Table \ref{tab6}.  Also, the exact analytical expressions for mass ratios ($\alpha, \beta$) have been provided in Table \ref{tab5}. With the help of some approximate analytical relations of neutrino mass ratios, we have checked the viability of all the six  with Tr $M_{\nu}$=0 condition.
\subsection{ CP violating textures ($P_{1}$, $P_{2}$, $P_{3}$)}

\subsubsection{Texture $P_{1}$ with  Tr $M_{\nu}=0$}\label{eq52}
Using Eqs. (\ref{eq28}) and (\ref{eq29}) and retaining only the leading order term  of $\theta_{13}$, we obtain  following analytical relations
\begin{equation}\label{eq52}
\alpha \equiv \dfrac{m_{1}}{m_{3}}\approx \dfrac{1}{\eta}\; sec2 \theta_{12} s_{12}^{2},
\end{equation}
\begin{equation}\label{eq53}
\beta \equiv \dfrac{m_{2}}{m_{3}}\approx -\dfrac{1}{\kappa}\; sec2 \theta_{12} c_{12}^{2},
\end{equation}
where $\eta= +1$, $\kappa=-1$ so that $\alpha, \beta$ remain real and positive. Using Eqs. (\ref{eq52}) and (\ref{eq53}), we obtain  $R_{\nu}\approx \beta^{2}-\alpha^{2} \approx sec2\theta_{12}$ for normal mass ordering.  Using 3$\sigma$ experimental range of oscillation parameter, we find,  $2.23\leq R_{\nu} \leq 4.02$, which excludes the experimental range of  $R_{\nu}$  and for inverted mass ordering , we have
\begin{equation}\label{eq54}
R_{\nu}\approx \dfrac{sec2 \theta_{12}}{sec^{2}2 \theta_{12} c^{4}_{12}-1},
\end{equation}
which is again inconsistent with current experimental data as $R_{\nu}> 0.75$. Therefore, texture $P_{1}$ with Tr $M_{\nu}=0$  is ruled out according to latest  neutrino oscillation data at 3$\sigma$ CL.\\
\subsubsection{Texture $P_{2}$ with  Tr $M_{\nu}=0$ }
\begin{figure}[h!]
\begin{center}

\subfigure[]{\includegraphics[width=0.4\columnwidth]{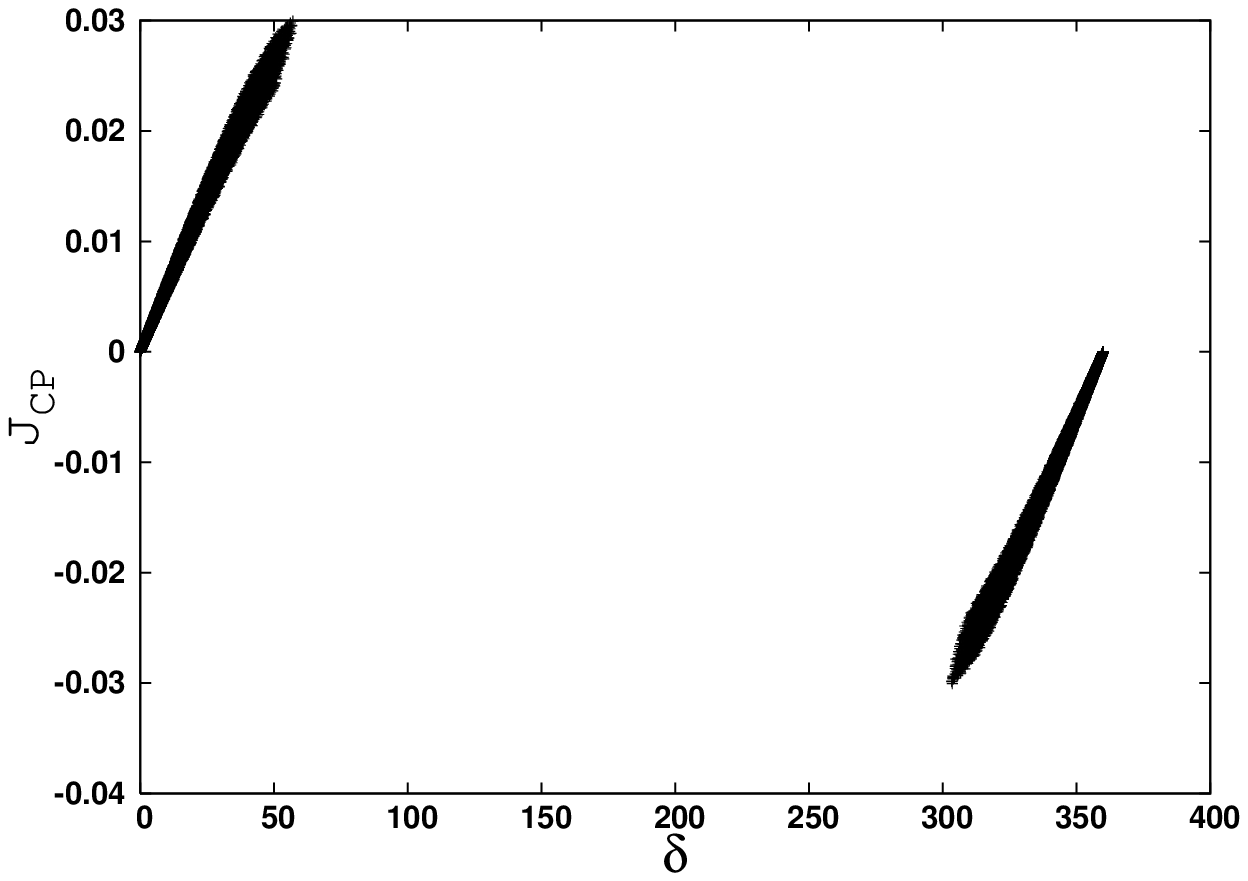}}
\subfigure[]{\includegraphics[width=0.4\columnwidth]{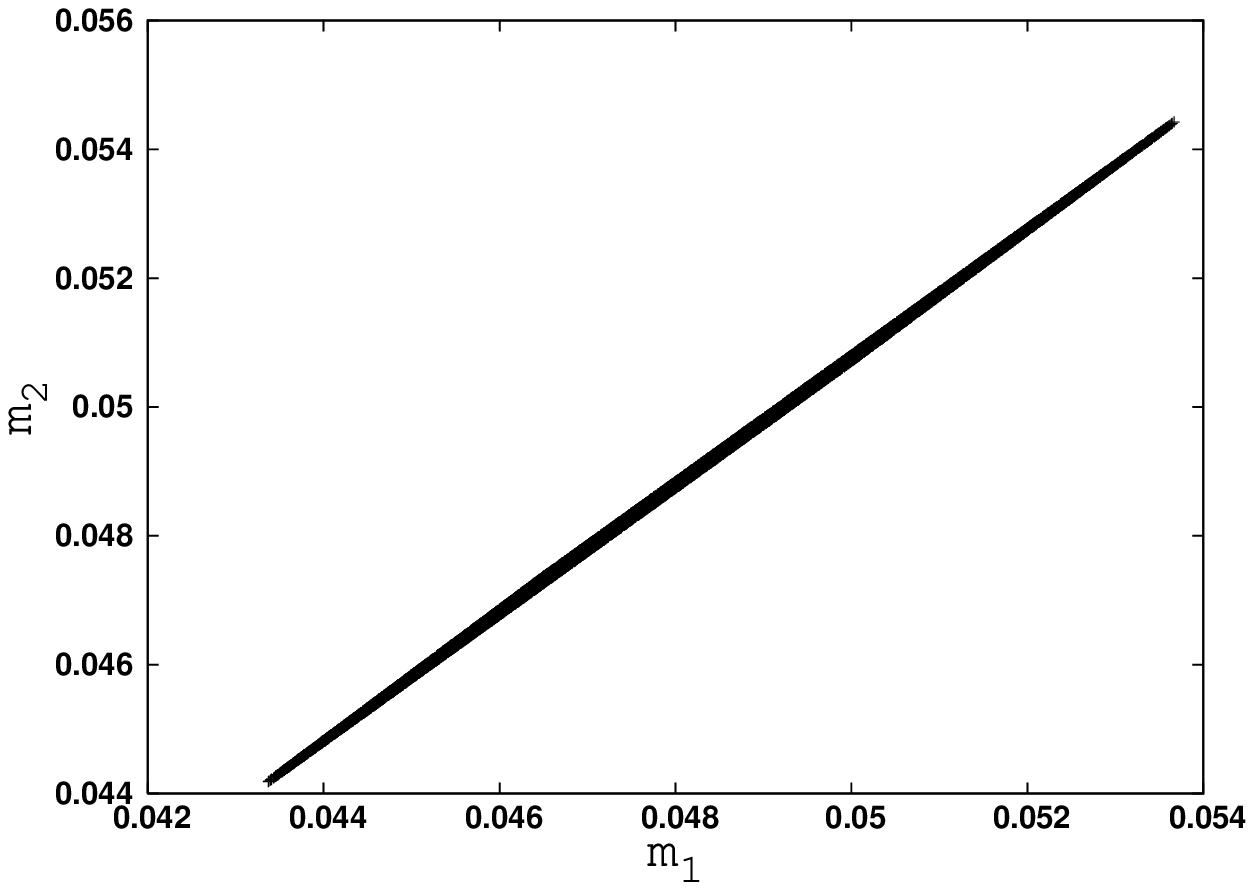}}
\subfigure[]{\includegraphics[width=0.4\columnwidth]{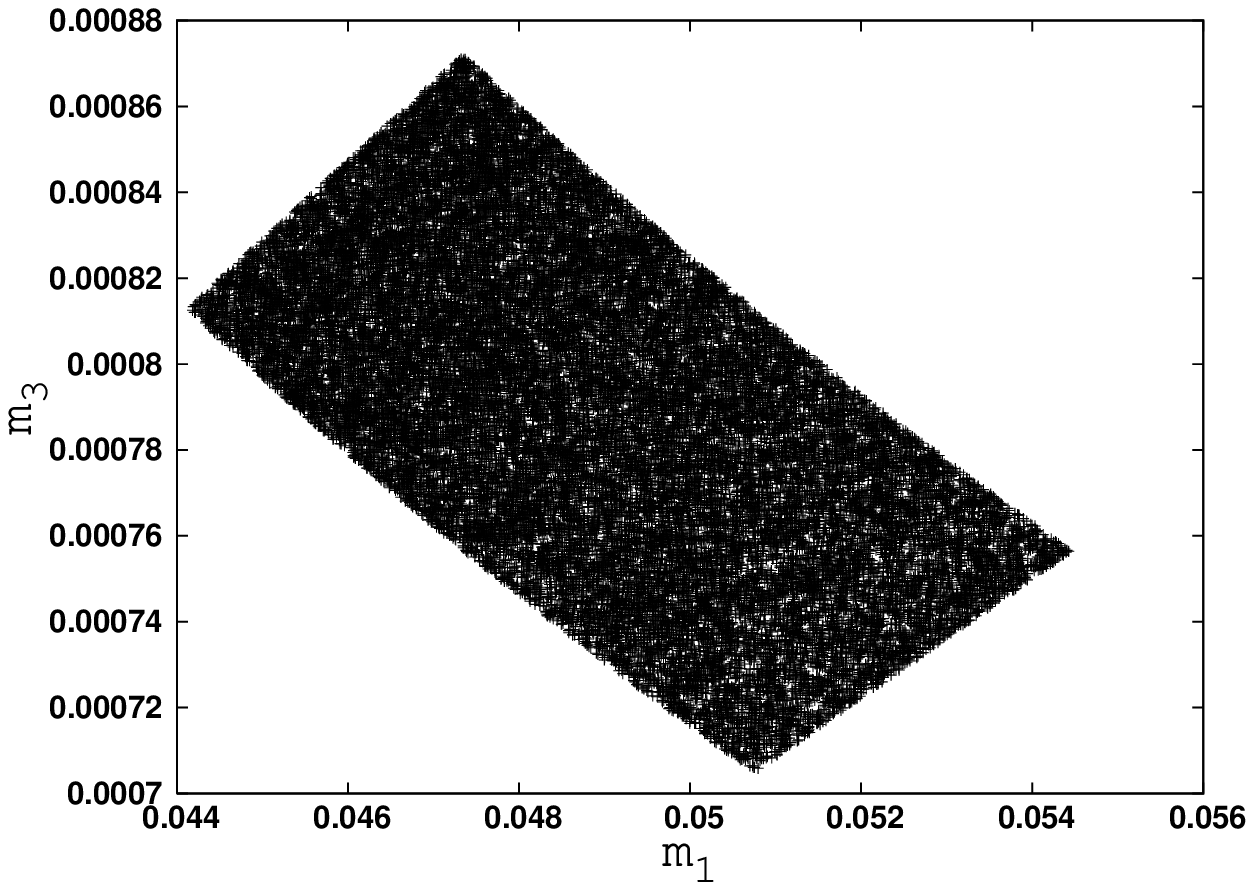}}\\
\caption{\label{fig4}  Correlation plots for  $P_{2}$ with Tr $M_{\nu}=0$. The neutrino masses $m_{1, 2, 3}$ are measured in eV. }
\end{center}
\end{figure}

Using Eqs. (\ref{eq28}) and (\ref{eq29}), we obtain the following analytical relations in the leading order approximation  of $\theta_{13}$
\begin{equation}\label{eq55}
\alpha\equiv\dfrac{m_{1}}{m_{3}}\approx -\dfrac{1}{\eta}\; sec2 \theta_{12}( c_{12}^{2}-t^{2}_{23}),
\end{equation}
\begin{equation}\label{eq56}
\beta \equiv \dfrac{m_{2}}{m_{3}}\approx \dfrac{1}{\kappa}\; sec2 \theta_{12} (s_{12}^{2}-t^{2}_{23}).
\end{equation}
Here, $\eta=+1, \kappa=-1$ must hold so as to get non-negative mass ratios $\frac{m_{1}}{m_{3}}$ and $\frac{m_{2}}{m_{3}}$. From figure \ref{fig4}(a), it is evident that parameter space of Dirac CP violating phase lies in
the range,  $\delta \in [0^{0},56^{0}]\oplus[306^{0},360^{0}]$. The
parameter space of ($J_{CP},\delta$) shows that $J_{CP}\neq0$,
implying  CP violation [figure \ref{fig4}(a)]. In figure \ref{fig4}(b, c), it is shown that only inverted mass ordering ($m_{3}<< m_{1}< m_{2}$) is allowed for this texture, while normal mass ordering is ruled out.
\subsubsection{Texture $P_{3}$ with  Tr $M_{\nu}=0$ }
\begin{figure}[h!]
\begin{center}

\subfigure[]{\includegraphics[width=0.4\columnwidth]{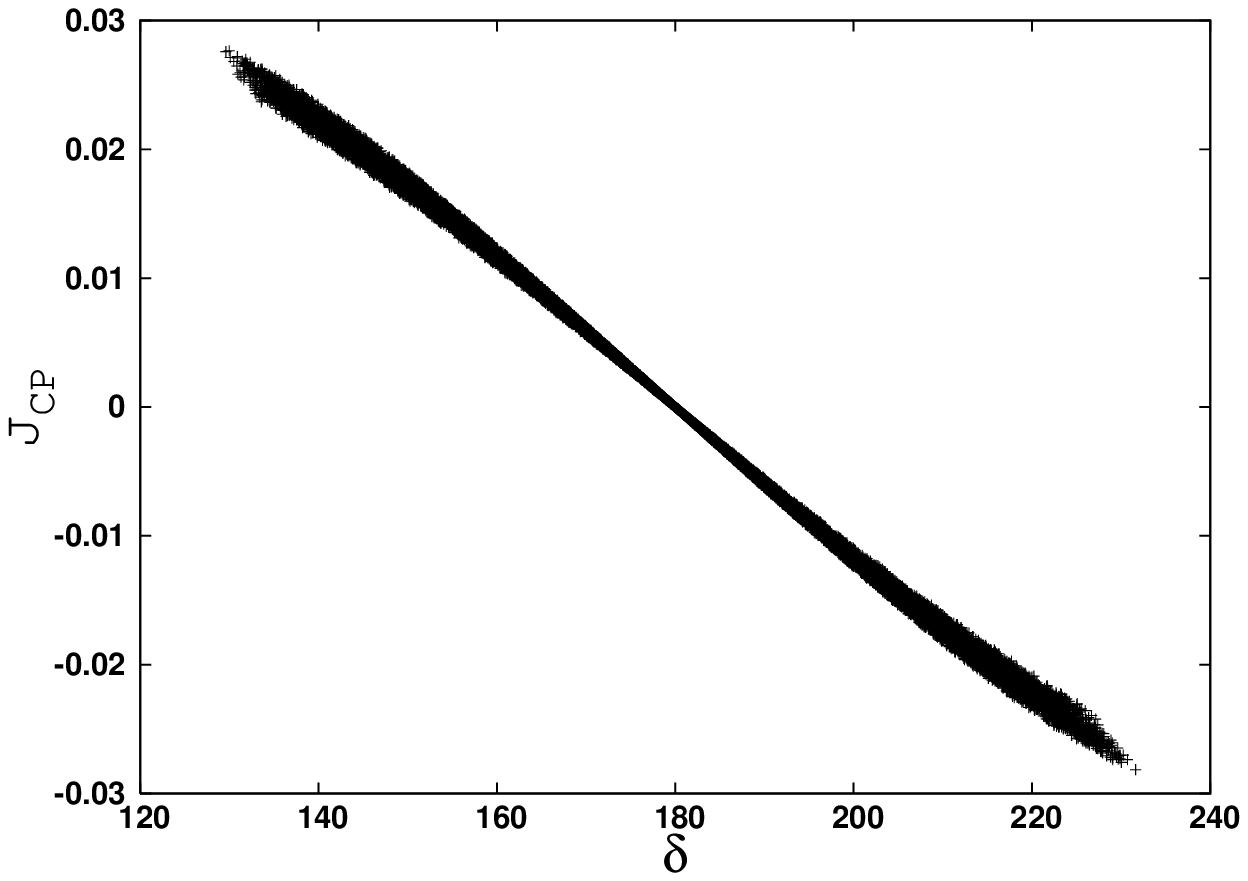}}
\subfigure[]{\includegraphics[width=0.4\columnwidth]{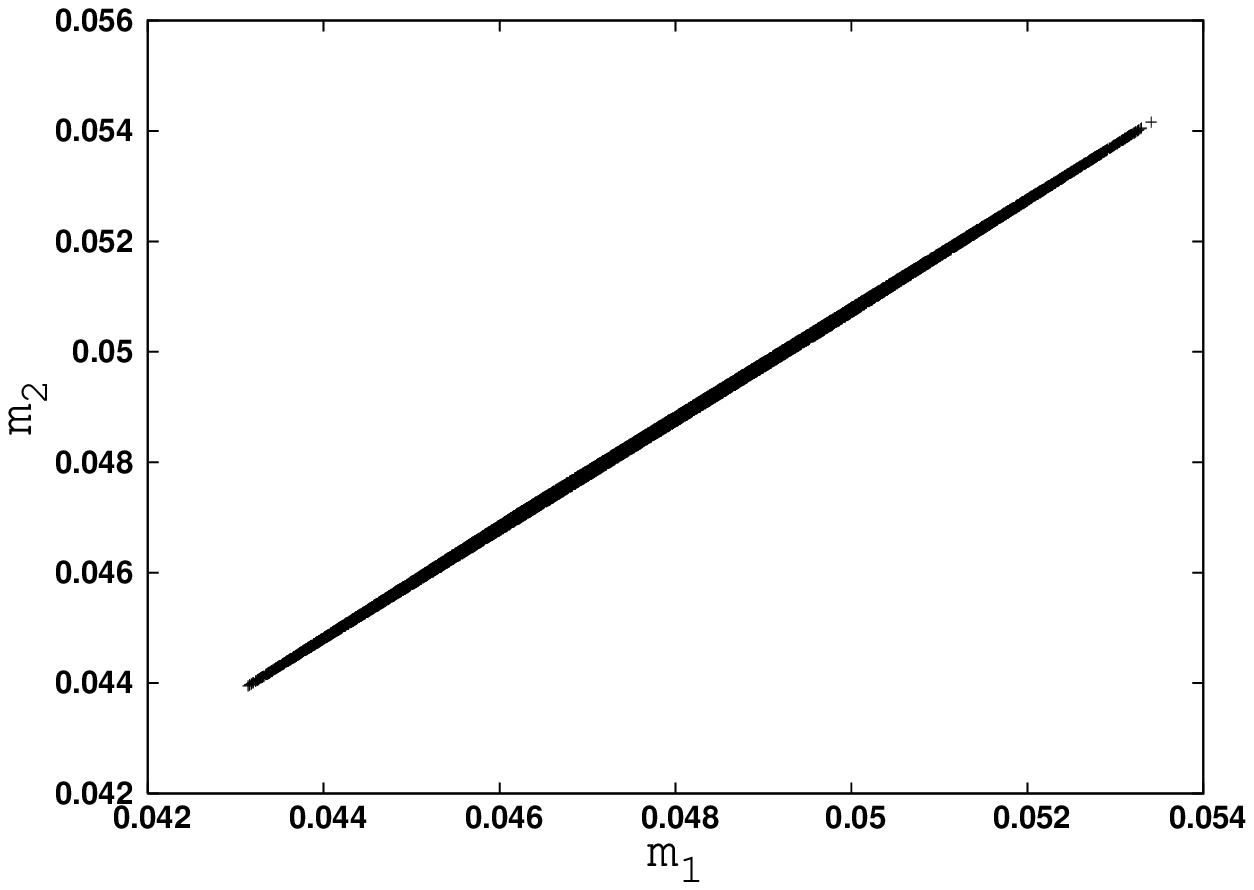}}
\subfigure[]{\includegraphics[width=0.4\columnwidth]{trp2m1m3.eps}}\\
\caption{\label{fig5} Correlation plots for $P_{3}$ with Tr $M_{\nu}=0$. The neutrino masses $m_{1, 2, 3}$ are measured in eV. }
\end{center}
\end{figure}

\begin{figure}[h!]
\begin{center}
\subfigure[]{\includegraphics[width=0.4\columnwidth]{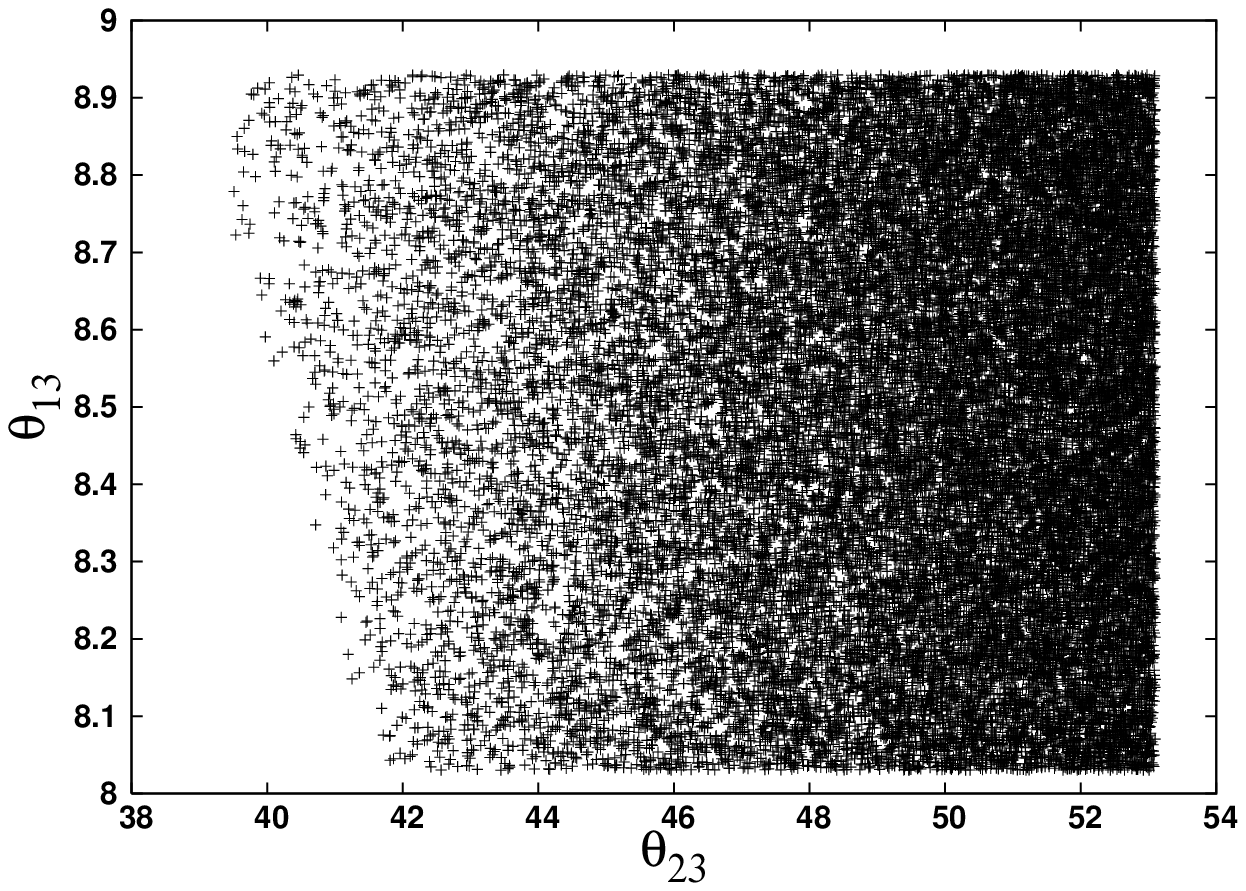}}
\subfigure[]{\includegraphics[width=0.4\columnwidth]{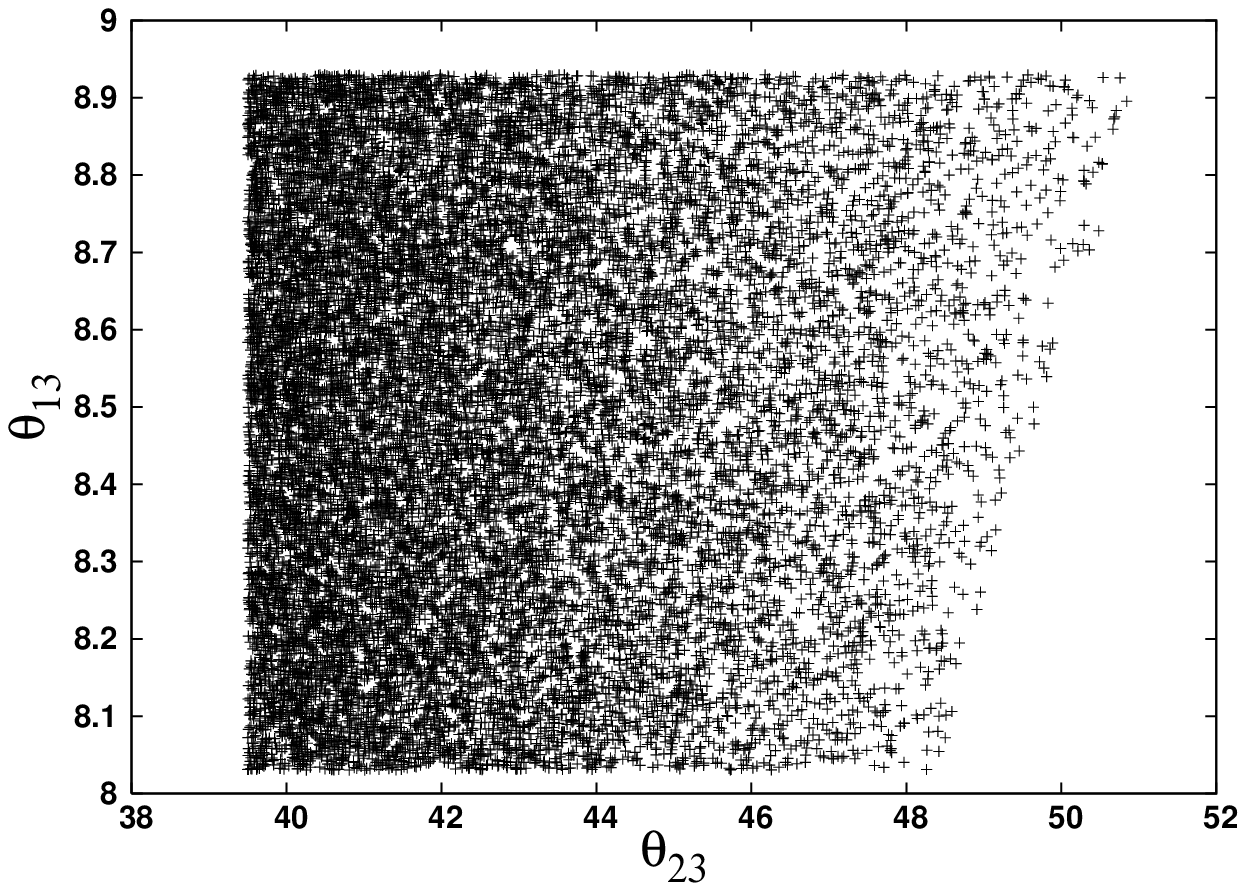}}\\
\caption{\label{fig6}  Correlation plots for textures (a) $P_{2}$  and (b) $P_{3}$,  depicting the permutation symmetry. }
\end{center}
\end{figure}

Since texture $P_3$ is related to texture $P_2$ via permutation symmetry as mentioned earlier, the phenomenological implications for texture $P_3$ can be obtained from texture $P_2$.  With the help of Eqs. (\ref{eq28}) and (\ref{eq29}), we deduce some useful analytical relations in the leading order term of $s_{13}$ term.
\begin{equation}\label{eq57}
\alpha\equiv\dfrac{m_{1}}{m_{3}}\approx -\dfrac{1}{\eta}\;sec2\theta_{12} \bigg(c^{2}_{12}-\frac{1}{t_{23}^{2}}\bigg),
\end{equation}
\begin{equation}\label{eq58}
\beta \equiv \dfrac{m_{2}}{m_{3}}\approx \dfrac{1}{\kappa}\;sec2\theta_{12} \bigg(s^{2}_{12}-\frac{1}{t_{23}^{2}}\bigg).
\end{equation}
Here, $\eta=+1, \kappa=-1$ must hold in order to obtain non-negative $\dfrac{m_{1}}{m_{3}}$ and $\dfrac{m_{2}}{m_{3}}$.  From figure \ref{fig4}(a), it is evident that parameter space of Dirac CP phase lies in the range,  $\delta \in [128.5^{0},231.5^{0}]$. The correlation plots for neutrino masses ($m_{1}, m_{2}, m_{3})$ indicates that only inverted mass ordering is allowed  [figures \ref{fig5} (b, c)]. In figure \ref{fig5} (b), there exist a strong linear correlation between neutrino masses $m_{1}$ and $m_{2}$.  The parameter space of ($J_{CP},\delta$) indicates $J_{CP}\neq0$, implying that texture $P_{3}$ exhibits  CP violation [figure \ref{fig5}(a)].
In figures.\ref{fig6}(a) and \ref{fig6}(b), we have provided the correlation plots between $\theta_{23}$ and $\theta_{13}$ for textures $P_{2}$ and $P_{3}$, respectively. Unlike Det $M_{\nu}=$0 case, textures $P_{2}$ and $P_{3}$ with Tr $M_{\nu}=0 $ prefer both the octant of $\theta_{23}$. Again, permutation symmetry between $P_{2}$ and $P_{3}$ is clearly visible in figure \ref{fig6}(a, b).

\subsection{CP conserving textures  ($P_{4}$, $P_{5}$, $P_{6}$)}
\subsubsection{Texture $P_{4}$ with Tr $M_{\nu}=0$}
Using Eqs. (\ref{eq28}) and (\ref{eq29}), we obtain the following analytical expressions in leading order of $s_{13}$ term
\begin{equation}\label{eq59}
\alpha\equiv \dfrac{m_{1}}{m_{3}}\approx -\dfrac{1}{\eta}\; 0.5,
\end{equation}
\begin{equation}\label{eq60}
\beta \equiv \dfrac{m_{2}}{m_{3}}\approx -\dfrac{1}{\kappa}\; 0.5.
\end{equation}
Since $R_{\nu}$=0 in the leading order approximation of $s_{13}$, we have to work at next to leading order of $\theta_{13}$
\begin{equation}\label{eq61}
\alpha \equiv \dfrac{m_{1}}{m_{3}}\approx -\dfrac{1}{\eta}\;\dfrac{1}{2}\bigg(1\mp \dfrac{3}{2} \dfrac{s_{23} s_{13}}{c_{12}s_{12}c_{23}}\bigg)+O(s^{2}_{13}),
\end{equation}
\begin{equation}\label{eq62}
\beta \equiv \dfrac{m_{2}}{m_{3}}\approx -\dfrac{1}{\kappa}\;\dfrac{1}{2}\bigg(1\pm \dfrac{3}{2} \dfrac{s_{23} s_{13}}{c_{12}s_{12}c_{23}}\bigg)+O(s^{2}_{13}).
\end{equation}
Here $\eta. \kappa= \pm1$ and the upper and lower signs in the above expression refer to cases $\delta=0^{0}$ and $180^{0}$ , respectively. For $\delta= 180^{0}$ ,we find $m_{2}<m_{1}$, which is in contradiction with solar neutrino oscillation data. Using Eqs. (\ref{eq61}) and (\ref{eq62}), $R_{\nu}$ can be given as
\begin{equation}\label{eq63}
R_{\nu}\approx \beta^{2}-\alpha^{2} \approx \dfrac{3}{2}\bigg(\dfrac{s_{23} s_{13}}{c_{12}s_{12}c_{23}}\bigg).
\end{equation}
For normal mass ordering, taking into account the 3$\sigma$ experimental range of oscillation parameters, we find, 0.235$\leq R_{\nu} \leq$ 0.468, which excludes the experimental range of  $R_{\nu}$. Similarly, for inverted mass ordering, texture $P_{4}$ is found to be ruled out. Since textures $P_{5}$ and $P_{4}$ are phenomenologically related to each other due to permutation symmetry, therefore texture $P_{5}$ also remains inconsistent with latest experimental data at 3$\sigma$ level.

\subsubsection{Texture $P_{6}$ with Tr $M_{\nu}=0$}
Using Eqs. (\ref{eq28}) and (\ref{eq29}), we deduce some analytical expressions in leading order of $s_{13}$ term.
\begin{equation}\label{eq64}
\alpha\equiv \dfrac{m_{1}}{m_{3}}\approx -\dfrac{1}{\eta}\; sec2 \theta_{12} (1+c_{12}^{2}),
\end{equation}
\begin{equation}\label{eq65}
\beta \equiv \dfrac{m_{2}}{m_{3}}\approx \dfrac{1}{\kappa}\; sec2 \theta_{12} (1+s_{12}^{2}),
\end{equation}
where $\eta. \kappa=\pm1$. From  Eqs. (\ref{eq64}) and (\ref{eq65}), we find that $m_{2}<m_{1}$, therefore, texture $P_{6}$  with Tr $M_{\nu}=0$  is ruled out for both normal as well as inverted mass ordering at 3$\sigma$ CL.

The neutrino mass matrices for two allowed textures  viz. $P_{2}$ and $P_{3}$ with Tr $M_{\nu}=0$ are given below:\\
\begin{equation*}
M_{\nu}^{P_{2}}=
\begin{pmatrix}
c_{11} & c_{12} & c_{13}\\
c_{21} & c_{22} & c_{23}\\
c_{31} & c_{32} & c_{33}
\end{pmatrix},
\end{equation*}
where
\begin{align*}
&c_{11}=0.0104-0.0211,\\
&c_{12}=((-0.0402)-(-0.0154))+i((-0.0252)-0.0253),\\
&c_{13}=(0.0178-0.0374)+i((-0.0329)-0.0324),\\
&c_{21}=((-0.0402)-(-0.0154))-i((-0.0252)-0.0253), \\
&c_{22}=0.0,\\
&c_{23}=(0.00506-0.00926)+i((-0.00651)-0.00638),\\
&c_{31}=(0.0178-0.0374)-i((-0.0329)-0.0324),\\
&c_{32}=(0.00506-0.00926)-i((-0.00651)-0.00638),\\
&c_{33}=(-0.0211)-(-0.0104),
\end{align*}
and
\begin{equation*}
M_{\nu}^{P_{3}}=
\begin{pmatrix}
d_{11} & d_{12} & d_{13}\\
d_{21} & d_{22} & d_{23}\\
d_{31} & d_{32} & d_{33}
\end{pmatrix},
\end{equation*}
where
\begin{align*}
&d_{11}=0.0103-0.0193,\\
&d_{12}=(0.0196- 0.0369)+i(-0.0291-0.0297),\\
&d_{13}=((-0.0397)-(-0.0183))+i((-0.0293)-0.0297),\\
&d_{21}=(0.0196- 0.0369)-i((-0.0291)-0.0297) , \\
&d_{22}= (-0.0193)-(-0.0104),\\
&d_{23}=(0.00544-0.00912)+i((-0.00594)-0.00570),\\
&d_{31}=((-0.0397)-(-0.0183))-i((-0.0293)-0.0297),\\
&d_{32}=(0.00544-0.00912)-i((-0.00594)-0.00570),\\
&d_{33}=0.0.
\end{align*}

From above matrices, we observe that the elements of mass matrices for textures $P_{2}$ and $P_{3}$ are approximately similar to mass matrix elements for texture one zero with Det $M_{\nu}$=0 condition. For the sake of comparison, the range of $\delta$ has been provided for both the conditions  [Table \ref{tab7} ].\\

\section{Summary and conclusions}
To summarize our analysis, we have studied the implication of Det $M_{\nu}$=0 or Tr $M_{\nu}$=0 conditions, on texture one zero neutrino mass matrices, assuming the Dirac nature of neutrinos.
The six viable textures have been broadly classified into two categories viz. CP violating
($P_{1}, P_{2}, P_{3}$) and CP conserving ($P_{4}, P_{5}, P_{6}$), respectively. Therefore, CP violation is only possible for $P_{1,2,3}$, and we have $\delta = 0$ or $\pi$ for the other one-zero textures.
 In the analysis, all the CP conserving textures are found to be ruled out for both normal as well as inverted mass ordering at $3\sigma$ CL, however among the CP violating textures, only $P_{1}$ and $P_{2}$ are able to survive
the data for inverted mass ordering. 
 
In Ref.\cite{40,42}, it is explicitly shown that  CP-conserving value $\delta$ = 0 (or 2$\pi$) is disfavored at 3$\sigma$ in both NO and IO, while  $\delta = \pi$ is also disfavored at 3$\sigma$ in IO but not in NO (where it is still allowed at 2$\sigma$). In addition, a preference for CP violation with sin$\delta< 0$ is indicated at $<2\sigma$ CL \cite{37,39,40,42}. These experimental indications  are motivating as far as our analysis is concerned, however, a precise determination of $\delta$  is important to decide the compatibility of these textures. In the end, the phenomenological results of survived textures have been compared for both the conditions.

To conclude our discussion, we would like to mention that it is very difficult to determine
the exact nature of neutrinos whether Dirac or Majorana particle under the current experimental scenario. Therefore the assumption of Dirac neutrino carries some motivation. The
only possibility in the near future depends on neutrinoless double beta decay experiments,
which would determine the Majorana nature of neutrinos. In addition, the absolute neutrino
mass is still not known, therefore the consideration of vanishing neutrino mass or vanishing sum of neutrino masses can not be ruled out at present. The data collected from the
Planck satellite \cite{43} combined with other cosmological data, however put a upper limit on the sum of neutrino masses as  $\sum_{i}m_{i}<0.23 ~eV$ at 2$\sigma$ CL. The future and currently running longbaseline experiments, neutrinoless double beta decay experiments and cosmological
observations would check the validity of the present analysis and assumptions.

\section*{Acknowledgment}

The author would like to thank the Director, National Institute of technology (NIT) Kurukshetra, India for providing neccessary facilities to work, and special thanks to Dr. Radha Raman Gautam for the useful discussion during this work.\\

\begin{table}
\begin{center}
\begin{tabular}{|c|c|c|}
  \hline
  &&\\
  Texture & $\dfrac{m_{2}} {m_{3}}$ & status \\
   &&\\
 \hline
 &&\\
  $P_{1}$ &$ -\dfrac{1}{\kappa}\;\dfrac{t_{13}^{2}} {s_{12}^{2}}$  & unviable \\
 &&\\
  \hline
   &&\\
  $P_{2}$ & $-\dfrac{1}{\kappa}\;\dfrac{s_{23}^{2} c_{13}^{2}}{s_{12}^{2}s_{23}^{2}s_{13}^{2}+c_{12}^{2}c_{23}^{2}-2s_{12}s_{23}s_{13}c_{12}c_{23}c_{\delta}}$ & unviable \\
 &&\\
  \hline
   &&\\
   $P_{3}$& $-\dfrac{1}{\kappa}\;\dfrac{c_{23}^{2} c_{13}^{2}}{s_{12}^{2}c_{23}^{2}s_{13}^{2}+c_{12}^{2}s_{23}^{2}+2s_{12}s_{23}c_{12}c_{23}s_{13}c_{\delta}}$ & unviable \\
    &&\\
   \hline
   &&\\
  $P_{4}$ &$-\dfrac{1}{\kappa}\;\dfrac{s_{23}s_{13}} {-s_{12}^{2} s_{23}s_{13}\pm s_{12}c_{12}c_{23}}$ & unviable \\
   &&\\
  \hline
   &&\\
  $P_{5}$ &$ -\dfrac{1}{\kappa}\;\dfrac{c_{23}c_{13}s_{13}} {-s_{12}^{2}c_{23}c_{13}s_{13}\mp s_{12}c_{12}c_{13}s_{23}}$ & unviable \\
   &&\\
  \hline
    &&\\
  $P_{6}$ & $-\dfrac{1}{\kappa}\;\dfrac{s_{23} c_{23} c_{13}^{2}}{(s_{12}^{2}s_{13}^{2}-c_{12}^{2})c_{23}s_{23}+s_{12}c_{12}s_{13}(\pm s_{23}^{2} \mp c_{23}^{2})}$ & unviable \\
   &&\\
   \hline
    
\end{tabular}
\caption{\label{tab3}The exact expressions of mass ratio $\dfrac{m_{2}} {m_{3}}$ along with the status of all the six one zero textures with $m_{1}=0$ (normal mass ordering) is shown. The upper and lower signs in the above expressions refer to cases $\delta=0^{0}$ and $180^{0}$, respectively.}
\end{center}
\end{table}

\begin{table}
\begin{center}
\begin{tabular}{|c|c|c|}
  \hline
  &&\\
  Texture &$\dfrac{m_{2}} {m_{1}}$  & status \\

  \hline
  &&\\
  $P_{1}$ &$-\dfrac{\eta}{\kappa}\;\dfrac{1} {t_{12}^{2}}$  & unviable \\
&&\\
  \hline
   &&\\
  $P_{2}$ & $-\dfrac{\eta}{\kappa}\;\dfrac{c_{12}^{2}s_{23}^{2}s_{13}^{2}+s_{12}^{2}c_{23}^{2}+2s_{12}s_{23}c_{12}c_{23}s_{13}c_{\delta}}{s_{12}^{2}s_{23}^{2}s_{13}^{2}+c_{12}^{2}c_{23}^{2}-2s_{12}s_{23}c_{12}c_{23}s_{13}c_{\delta}}$ & viable \\
   &&\\
   \hline
    &&\\
   $P_{3}$& $-\dfrac{\eta}{\kappa}\;\dfrac {c_{12}^{2}c_{23}^{2}s_{13}^{2}+s_{12}^{2}s_{23}^{2}-2s_{12}s_{23}c_{12}c_{23}s_{13}c_{\delta}}{s_{12}^{2}c_{23}^{2}s_{13}^{2}+c_{12}^{2}s_{23}^{2}+2s_{12}s_{23}c_{12}c_{23}s_{13}c_{\delta}}$ & viable \\
    &&\\
   \hline
    &&\\
    $P_{4}$ &$-\dfrac{\eta}{\kappa}\;\dfrac{1}{t_{12}}.\dfrac{-s_{23}c_{12}s_{13}\mp s_{12}c_{23}}{-s_{23}s_{12}s_{13} \pm c_{12}c_{23}}$ & unviable \\
   &&\\
  \hline
   &&\\
  $P_{5}$ &$-\dfrac{\eta}{\kappa}\;\dfrac{1}{t_{12}}.\dfrac{-c_{12}c_{23}s_{13}\pm s_{12}s_{23}}{-s_{12}c_{23}s_{13} \mp c_{12}s_{23}}$ &unviable \\
   &&\\
  \hline
  $P_{6}$ & $-\dfrac{\eta}{\kappa}\;\dfrac{s_{23}c_{23}(c_{12}^{2}s_{13}^{2}-s_{12}^{2})+s_{12}c_{12}s_{13}(\pm c_{23}^{2} \mp s_{23}^{2})}{s_{23}c_{23}(s_{12}^{2}s_{13}^{2}-c_{12}^{2})+s_{12}c_{12}s_{13}(\pm s_{23}^{2} \mp c_{23}^{2})}$ & unviable \\
   &&\\
  \hline
   
  \end{tabular}
   \caption{\label{tab4}The expressions of mass ratio $\dfrac{m_{2}} {m_{1}}$ alongwith the status of all the six one zero textures with $m_{3}=0$(inverted mass ordering) is shown. The upper and lower signs in the above expressions refer to cases $\delta=0^{0}$ and $180^{0}$, respectively.}
\end{center}
\end{table}

\begin{table}
\begin{center}
\begin{tabular}{|c|c|c|c|}
  \hline
  &&&\\
  
  Texture & Analytical expressions for $\dfrac{m_{1}}{m_{3}}$ and $\dfrac{m_{2}}{m_{3}}$&NO&IO  \\

  \hline
  &&&\\
  $P_{1}$ & $\alpha=\dfrac{1}{\eta}\; sec 2\theta_{12}(s_{12}^{2}- t_{13}^{2})$ &&\\
& $\beta =-\dfrac{1}{\kappa}\; sec 2\theta_{12}(c_{12}^{2}- t_{13}^{2})$& unviable&unviable\\
&&&\\
  \hline
  &&&\\
  $P_{2}$ &  $\alpha=\dfrac{1}{\eta}\;\dfrac{(s_{12}^{2}s_{13}^{2}-c_{13}^{2})s_{23}^{2}+c_{12}c_{23}(c_{12}c_{23}-2s_{12}s_{23}s_{13}c_{\delta})} {(s_{23}^{2}s_{13}^{2}-c_{23}^{2})c_{2(12)}+s_{2(12)}s_{2(23)}s_{13} c_{\delta}}$ &&\\
  &$\beta=\dfrac{1}{\kappa}\;\dfrac{(-c_{12}^{2}s_{13}^{2}+c_{13}^{2})s_{23}^{2}-s_{12}c_{23}(s_{12}c_{23}+2c_{12}s_{23}s_{13}c_{\delta})} {(s_{23}^{2}s_{13}^{2}-c_{23}^{2})c_{2(12)}+s_{2(12)}s_{2(23)}s_{13} c_{\delta}}$& unviable&viable\\
  &&&\\
   \hline
   &&&\\
  $P_{3}$ &  $\alpha=\dfrac{1}{\eta}\;\dfrac{(s_{12}^{2}s_{13}^{2}-c_{13}^{2})c_{23}^{2}+c_{12}s_{23}(c_{12}s_{23}+2s_{12}c_{23}s_{13}c_{\delta})} {(c_{23}^{2}s_{13}^{2}-s_{23}^{2})c_{2(12)}-s_{2(12)}s_{2(23)}s_{13} c_{\delta}}$&& \\
  &$\beta=\dfrac{1}{\kappa}\;\dfrac{(-c_{12}^{2}s_{13}^{2}+c_{13}^{2})c_{23}^{2}-s_{12}s_{23}(s_{12}s_{23}-2c_{12}c_{23}s_{13}c_{\delta})} {(c_{23}^{2}s_{13}^{2}-s_{23}^{2})c_{2(12)}-s_{2(12)}s_{2(23)}s_{13} c_{\delta}}$& unviable&viable\\
  &&&\\
   \hline
  
  &&&\\
  $P_{4}$ &  $\alpha=\dfrac{1}{\eta}\;\dfrac{s_{23}s_{13}(1+s_{12}^{2})\mp s_{12}c_{12}c_{23}}{s_{23}s_{13}(c_{12}^{2}-s_{12}^{2})\pm 2s_{12}c_{12}c_{23}}$&&\\
   &  $\beta=-\dfrac{1}{\kappa}\;\dfrac{s_{23}s_{13}(1+c_{12}^{2})\pm s_{12}c_{12}c_{23}}{s_{23}s_{13}(c_{12}^{2}-s_{12}^{2})\pm 2s_{12}c_{12}c_{23}}$& unviable&unviable\\
   &&&\\
  \hline
  &&&\\
  $P_{5}$ &  $\alpha=\dfrac{1}{\eta}\;\dfrac{c_{23}s_{13}(1+s_{12}^{2})\pm s_{12}c_{12}c_{23}}{c_{23}s_{13}(c_{12}^{2}-s_{12}^{2})\mp 2s_{12}c_{12}s_{23}}$&&\\
   &  $\beta=-\dfrac{1}{\kappa}\;\dfrac{c_{23}s_{13}(1+c_{12}^{2})\mp s_{12}c_{12}s_{23}}{c_{23}s_{13}(c_{12}^{2}-s_{12}^{2})\mp 2s_{12}c_{12}s_{23}}$& unviable&unviable\\
   &&&\\
  \hline
  &&&\\
   $P_{6}$ &  $\alpha=\dfrac{1}{\eta}\;\dfrac{s_{23}c_{23}(s_{12}^{2}s_{13}^{2}-c_{12}^{2}-c_{13}^{2})+c_{12}s_{12}s_{13}(\pm s_{23}^{2}\mp c_{23}^{2})}{s_{23}c_{23}(1+s_{13}^{2})c_{2(12)}\mp 2s_{12}c_{12}s_{13}}$&& \\
   &  $\beta=\dfrac{1}{\kappa}\;\dfrac{s_{23}c_{23}(-c_{12}^{2}s_{13}^{2}+s_{12}^{2}+c_{13}^{2})+c_{12}s_{12}s_{13}(\mp s_{23}^{2}\pm c_{23}^{2})}{s_{23}c_{23}(1+s_{13}^{2})c_{2(12)}\mp 2s_{12}c_{12}s_{13}}$& unviable&unviable \\
  &&&\\  
 \hline
  \end{tabular}
   \caption{ \label{tab5} The exact expressions of mass ratios $\alpha \equiv \dfrac{m_{1}} {m_{3}}$ and $\beta \equiv \dfrac{m_{2}} {m_{3}}$ alongwith the status of all the six one zero textures with vanishing trace (i.e. Tr $M_{\nu}=0$) is shown. The upper and lower signs in the above expressions refer to the cases $\delta=0^{0}$ and $180^{0}$ , respectively. The symbols $c_{2(ij)}$ $\equiv$ cos 2$\theta_{ij}$, $s_{2(ij)}$ $\equiv$ sin 2$\theta_{ij}$ are defined.}
\end{center}
\end{table}
\newpage

\begin{table}
\begin{center}
\begin{tabular}{|c|c|c|c|}
  \hline
  
  Texture & signs of $\eta$ and $\kappa$ &NO&IO  \\

  \hline
  
  $P_{1}$ & $\eta=+1, \kappa =+1$ &$\times$& $\times$\\

& $\eta=+1, \kappa =-1$ &$\times$& $\times$\\
& $\eta=-1, \kappa =+1$ &$\times$& $\times$\\
& $\eta=-1, \kappa =-1$ &$\times$& $\times$\\
\hline
  
  $P_{2}$ & $\eta=+1, \kappa =+1$ &$\times$& $\times$\\

& $\eta=+1, \kappa =-1$ &$\times$& allowed\\
& $\eta=-1, \kappa =+1$ &$\times$& $\times$\\
& $\eta=-1, \kappa =-1$ &$\times$& $\times$\\

   \hline
   
  $P_{3}$ & $\eta=+1, \kappa =+1$ &$\times$& $\times$\\

& $\eta=+1, \kappa =-1$ &$\times$& allowed\\
& $\eta=-1, \kappa =+1$ &$\times$& $\times$\\
& $\eta=-1, \kappa =-1$ &$\times$& $\times$\\

   \hline
   
  $P_{4}$ & $\eta=+1, \kappa =+1$ &$\times$& $\times$\\

& $\eta=+1, \kappa =-1$ &$\times$& $\times$\\
& $\eta=-1, \kappa =+1$ &$\times$& $\times$\\
& $\eta=-1, \kappa =-1$ &$\times$& $\times$\\

  \hline
  
  $P_{5}$ & $\eta=+1, \kappa =+1$ &$\times$& $\times$\\

& $\eta=+1, \kappa =-1$ &$\times$& $\times$\\
& $\eta=-1, \kappa =+1$ &$\times$& $\times$\\
& $\eta=-1, \kappa =-1$ &$\times$& $\times$\\
  
  \hline
 
  $P_{6}$ & $\eta=+1, \kappa =+1$ &$\times$& $\times$\\

& $\eta=+1, \kappa =-1$ &$\times$& $\times$\\
& $\eta=-1, \kappa =+1$ &$\times$& $\times$\\
& $\eta=-1, \kappa =-1$ &$\times$& $\times$\\
 
  \hline
  \end{tabular}
   \caption{ \label{tab6} All possibilities of signs of $\eta$ and $\kappa$, which are associated with the expressions of mass ratios $\frac{m_{1}} {m_{3}}$ and $\frac{m_{2}} {m_{3}}$ alongwith the status of all the six one zero textures with Det $M_{\nu}=0$ or  Tr $M_{\nu}=0$) is shown. }
\end{center}
\end{table}
\newpage

\begin{table}
\begin{center}
\begin{tabular}{|c|c|c|}
  \hline
  Texture &$(M_{\nu})_{lm}=0 $ with Det $M_{\nu}=0$  & $(M_{\nu})_{lm}=0 $ with Tr $M_{\nu}=0$  \\
\hline
  
  $P_{1}$ & $\times$ & $\times$  \\

  \hline
   
  $P_{2}$ & $\delta=0^{0}-53^{0}\oplus 306^{0}-360^{0}$ & $\delta=0^{0}-53^{0}\oplus 306^{0}-360^{0}$ \\
  &$J_{CP}=-0.0306-0.0300$&$J_{CP}=-0.0306-0.0300$\\
   \hline
   $P_{3}$ & $\delta=130^{0}-230^{0}$ & $\delta=128.5^{0}-231.8^{0}$ \\
  &$J_{CP}=-0.0291-0.0285$&$J_{CP}=-0.0291-0.0285$\\ 
   
   \hline
    $P_{4}$ & $\times$ & $\times$  \\
   
  \hline
   $P_{5}$ & $\times$ & $\times$  \\
  \hline
   $P_{6}$ & $\times$ & $\times$  \\
  \hline
  \end{tabular}
   \caption{\label{tab7} Comparison for allowed ranges of Dirac CP-violating phase($\delta$) and Jarlskog rephrasing invariant parameter($J_{CP}$) for all six one-zero textures  with Det $M_{\nu}=0$ and Tr $M_{\nu}=0$ respectively, is shown at 3$\sigma$ level. }
\end{center}
\end{table}

\end{document}